\documentclass[10pt,journal]{IEEEtran2}
\usepackage{amsmath,amssymb,amsthm}
\usepackage[noadjust]{cite}
\usepackage{gensymb}

\usepackage[font=footnotesize]{caption}
\usepackage{subcaption}
\usepackage{xspace,arydshln}
\usepackage{graphicx,fancyhdr,epstopdf}
\usepackage[ruled]{algorithm2e}
\usepackage{enumitem}
\usepackage{authblk}
\usepackage{color}
\usepackage{hyperref}
\usepackage{soul}
\usepackage{etoolbox}

\usepackage[compact]{titlesec}
\titlespacing{\section}{0pt}{2ex}{1ex}
\titlespacing{\subsection}{0pt}{1ex}{0ex}
\titlespacing{\subsubsection}{0pt}{0.5ex}{0ex}
\usepackage{microtype}

\definecolor{lee}{rgb}{0,0.8,0}
\definecolor{dblue}{rgb}{0,0,0.8}

\input arraymac.tex

\begin{document}

\title{Channel Estimation with Reconfigurable Intelligent Surfaces – A General Framework}
\author[1]{A.~Lee Swindlehurst~\IEEEmembership{Fellow,~IEEE}}
\author[2]{Gui Zhou~\IEEEmembership{Student Member,~IEEE}}
\author[3]{Rang Liu~\IEEEmembership{Student Member,~IEEE}}
\author[2]{Cunhua Pan~\IEEEmembership{Member,~IEEE}}
\author[3]{Ming Li,~\IEEEmembership{Senior Member,~IEEE}}%
\affil[1]{Center for Pervasive Communications \& Computing, University of California, Irvine, USA}
\affil[2]{School of Electronic Engineering \& Computer Science, Queen Mary University of London, England}
\affil[3]{School of Information \& Communication Engineering, Dalian University of Technology, China}
\maketitle

\begin{abstract}
Optimally extracting the advantages available from reconfigurable intelligent surfaces (RISs) in wireless communications systems requires estimation of the channels to and from the RIS. The process of determining these channels is complicated by the fact that the RIS is typically composed of passive elements without any data processing capabilities, and thus the channels must be estimated indirectly by a non-colocated device, typically a controlling base station. In this article, we examine channel estimation for RIS-based systems from a fundamental viewpoint. We study various possible channel models and the identifiability of the models as a function of the available pilot data and behavior of the RIS during training. In particular, we will consider situations with and without line-of-sight propagation, single- and multiple-antenna configurations for the users and base station, correlated and sparse channel models, single-carrier and wideband OFDM scenarios, availability of direct links between the users and base station, exploitation of prior information, as well as a number of other special cases. We further conduct numerical comparisons of achievable performance for various channel models using the relevant Cram\'er-Rao bounds. 
\end{abstract}

\section{Introduction}

There has been an explosion of interest in the use of reconfigurable metasurfaces for wireless communication systems in the last few years. Such reconfigurable intelligent surfaces (RIS) provide tunable degrees-of-freedom for adjusting the propagation characteristics of problematic channels (e.g., sparse channels with frequent blockages) that make them a valuable resource for maintaining and enhancing the quality of service (QoS) for users (UEs) in the network. However, most techniques that exploit this ability require channel state information (CSI) to and from the elements of the RIS, which is a challenge since the number of RIS elements may be very large, and more importantly, they are usually constructed only as passive devices without active transceivers or computational resources. Consequently, channel estimation for RIS-based systems has been a subject of intense study.

Because the RIS is passive, the CSI must be estimated by devices -- most often a basestation (BS) or access point -- that are not co-located with the RIS. For example, training signals transmitted by the UEs are received by the BS after reflection from the RIS, and possibly also over a direct path to the BS, and these known signals are exploited for CSI estimation. In order to estimate the RIS-based channel components, the reflection coefficients of the RIS must be varied as well, at least during a portion of the training period. However, even with variable training from the UEs and RIS, the fact that the impact of the RIS is only indirectly viewed in the data means that the complete structure of the channel is not identifiable. In particular, while the {\em cascaded} or {\em composite} channel from the UEs to the BS can be determined, the individual components of the channel involving the RIS cannot. Fortunately, this is typically not a problem for designing beamforming algorithms at the BS or optimizing the RIS reflection properties, since ultimately the QoS only depends on the composite channel.

A large amount of published work on CSI estimation for RIS-based systems has appeared recently. Initially, this work focused on estimating {\em unstructured} models, where the channels are simply described using complex gains \cite{LS-mvue,2019beixiong,2019hang,2020changsheng,2020Nadeem,2020dai,2020Zhaorui,2020Beixiong,2020LiWei,2021LiWei,ZhengYZ21}. Such models are simple and lead to straightforward algorithms, but the required training overhead is very large and may render such approaches impractical. Methods for reducing the training overhead, for example based on grouping the RIS elements or exploiting the common BS-RIS channel among the users, have been proposed, but larger reductions are possible when the channels are sparse if parametric or geometric channel models are used instead \cite{2019Chen,2020tensor,2020Ziwei,2020Peilan,2020SongNoh,MaCC20,ZhouPRPS21,2021Jiguang,2021dai,2021Ardah,MaSAH21,ChenTZ21,LiuZG21,LiuZW21}. In these models, the channels are parameterized by the angles of arrival (AoAs), angles of departure (AoDs) and complex gains of each propagation path. As long as the number of multipaths is not large, then the total number of parameters to be estimated can be 1-2 orders of magnitude smaller than in the unstructured case, and the amount of training can be correspondingly reduced. On the other hand, geometric models require knowledge of the array calibration and RIS element responses, as well as the model order; errors in the modeling assumptions will degrade some of this advantage. In addition, we will see later that the algorithms for estimating the geometric channel parameters can in general be quite complex.

Many CSI estimation techniques have been proposed under a wide array of assumptions, from Rayleigh fading to line-of-sight (LoS) propagation, single- to multi-antenna configurations, single- and multi-carrier modulation, scenarios with and without a direct link between BS and UEs, and a variety of other special cases. In this paper, we take a systematic approach to the problem and organize the various approaches that have been proposed -- as well as some that have not -- under a common framework. In this way, the advantages and disadvantages of different assumptions and solution approaches become clearer, and avenues for future work are elucidated.

After stating our general assumptions and notational conventions in Section~\ref{sec:assump}, we begin with a discussion of CSI estimation for unstructured channels in Section~\ref{sec:unstruc}. We will first consider the narrowband single user MIMO case and the corresponding least-squares (LS) and minimum mean squared error (MMSE) solutions, and then we will examine extensions to the wideband and multi-user cases, as well as special cases involving a single antenna BS and UEs and methods for reducing the training overhead. Then in Section~\ref{sec:geometric} we focus on estimation of geometric channel models, and we follow the same format of beginning with the narrowband single user MIMO case and then considering the same generalizations and special cases as in the previous section. Numerical examples involving the Cram\'er-Rao bound (CRB) will be given in Section~\ref{sec:crb} to illustrate the main conclusions. Several additional topics will be briefly considered in Section~\ref{sec:other}, including the use of some active transceivers at the RIS, scenarios with more than one RIS, machine learning approaches, etc. Finally some conclusions and suggestions for future research are offered in Section~\ref{sec:conc}.

\section{General Assumptions and Notation}\label{sec:assump}

In this paper, we primarily consider scenarios with a single basestation (BS), a single RIS, and potentially multiple co-channel UEs. Various assumptions are made about the number of antennas at the BS and UEs, and the number of UEs that are active. We assume the BS and UEs employ fully digital rather than hybrid digital/analog architectures. We also assume a standard time-division duplex protocol in which pilot symbols transmitted by the UEs in the uplink are exploited by the BS to obtain a channel estimate, which is then used for downlink beamforming or multiplexing. This assumes reciprocal uplink and downlink channels between the BS, RIS and users, which in turn typically requires some type of RF transceiver calibration and RIS elements whose behavior is independent of the angle of incidence. Pilots could also be embedded in the downlink for channel estimation at the UEs, but this is similar to the uplink problem and thus is not explicitly considered. 

Matrices and vectors are denoted by boldface capital and lowercase letters, respectively. In some cases, the $k$-th column or row of a matrix $\A$ will be denoted by $\A_{:k}$ or $\A_{k:}$, respectively. The transpose, conjugate transpose, and conjugate are denoted by $(\cdot)^T, (\cdot)^H$, and $(\cdot)^*$, respectively. The Kronecker, Khatri-Rao, and Hadamard products of two matrices are indicated by $\Cbf=\Abf\otimes\Bbf$, $\Cbf=\Abf\diamond\Bbf$ and $\Cbf=\Abf\odot\Bbf$, respectively. An $N\times N$ identity matrix is represented as $\Ibf_N$, and $N\times 1$ vectors composed of all ones or zeros are denoted by $\mathbf{1}_N$ and $\zerobf_N$, respectively. 
A circular complex multivariate Gaussian distribution with mean $\mubf$ and covariance $\Rbf$ is denoted by $\cal{CN}(\mubf,\Rbf)$. The function $\text{vec}(\A)$ creates a vector from matrix $\A$ by stacking its columns. The function $\lfloor a \rfloor$ creates an integer from real number $a$ by truncating its decimal part, and $a \, \text{mod} \, b$ is the modulo operator that returns the integer remainder of $a/b$. A diagonal matrix with elements of vector $\cbf$ on the diagonal is indicated by $\text{diag}(\cbf)$, and a block diagonal matrix with block entries $\C_1,\C_2,\ldots$ is written $\text{blkdiag}\left([\C_1 \; \; \C_2 \; \ldots]\right)$.

The reflective properties of an RIS with $N$ elements is described by the $N\times N$ diagonal matrix $\Phibf = \{{\rm diag}(\phibf)\}$, where $\phibf=[\beta_1 e^{j\alpha_1} \; \cdots \; \beta_N e^{j\alpha_N}]^T$. There are a number of practical issues associated with $\phibf$ that are important for RIS performance optimization, such as the dependence of the gains $\beta$ on the phases $\alpha$, the fact that the phases are typically discrete and frequency dependent, etc. For the most part, these issues are not directly relevant to the generic channel estimation problem, which only requires that $\phibf$ be known and sufficiently controllable. However, certain simplifying assumptions about $\phibf$ are made below for performance analyses or purposes of illustration.

\section{Estimation of Unstructured Channel Models}\label{sec:unstruc}

We begin with models where the channel between individual network elements is described by a complex coefficient in the case of a narrowband single carrier signal, or a complex-valued impulse response for wideband transmission. Such unstructured or nonparametric channel models are appropriate for situations with rich multipath scattering (e.g., sub-6GHz systems), where it is difficult to describe the aggregate characteristics of the propagation environment. We initially focus on the narrowband single-user scenario, and then examine cases involving wideband signals or multiple users. As will become clear, the limiting factor with unstructured CSI estimation is the large training overhead that is required. Approaches for reducing the training overhead are discussed at the end of the section.

\subsection{Narrowband Single User MIMO}
\label{sec:suscmimo}

\begin{figure}
\centering
  \includegraphics[width=3.5in]{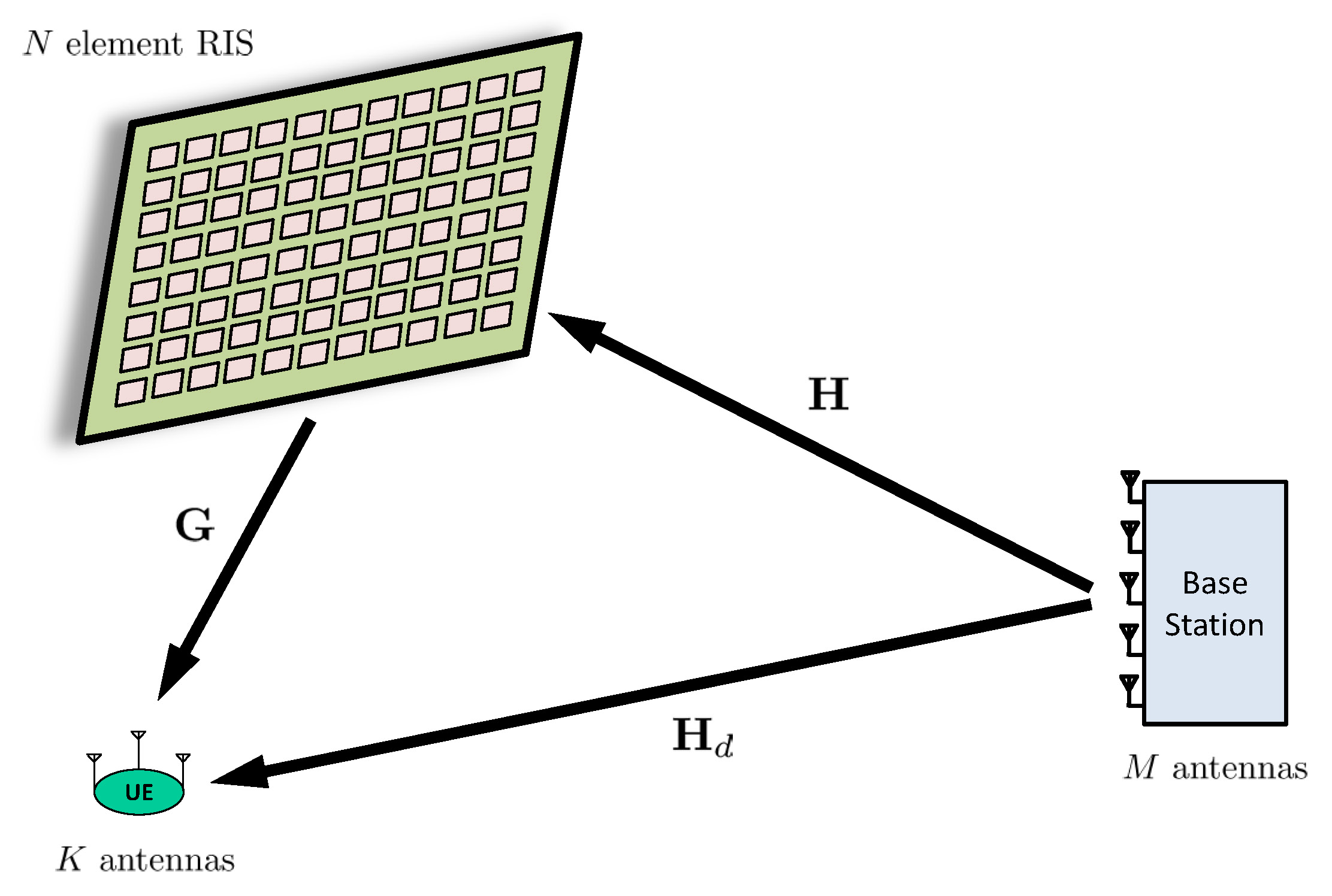} 
\caption{A generic scenario involving an $M$-antenna basestation, an $N$-element RIS, and a $K$-element user.}
\label{fig1}
\vskip -0.5cm
\end{figure}

The scenario assumed here is as depicted in Fig.~\ref{fig1}, with an $M$-antenna BS, an $N$-element RIS, and a single UE with $K$ antennas. The geometries of the RIS and the arrays at the BS and UE are arbitrary. If the UE transmits the $K\times 1$ vector $\xbf_t$ at time $t$, the signal received at the BS is given by 
\beq[sumimo]
\ybf_t = \sqrt{P} \left( \Hbf_d + \Hbf \Phibf_t \Gbf^H \right) \xbf_t + \nbf_t \; , 
\eeq
where ${\H_d,\H,\G}$ are respectively the channels between the BS and UE, the BS and RIS, and the RIS and UEs, and $\nbf_t$ denotes additive noise or interference. Assuming $\Ec\{\xbf_t \xbf_t^H\}=\I_K$ and $\nbf_t \sim \Cc\Nc(\zerobf,\sigma^2 \Ibf_M)$, $P$ represents the transmit power, and the signal-to-noise ratio (SNR) is defined as $P/\sigma^2$. The channels are all assumed to be block flat fading and constant over a coherence interval sufficiently long to permit channel estimation and subsequent data transmission. On the other hand, the reflection coefficients of the RIS, $\Phibf_t$, can vary synchronously with the UE uplink transmission. Some prior work ignores the direct channel component $\Hbf_d$, assuming that it is either not present (e.g., due to a blockage), or that it was estimated in a previous step and its contribution has been removed from the received data, {\em i.e.,} $\ybf_t \longrightarrow \ybf_t - (1/\sqrt{P}) \Hbf_d\xbf_t$. 

It is important to note that not all of the components of the channel-related term $\Hbf_d+\Hbf\Phibf_t\Gbf^H$ are individually identifiable. In particular, for any invertible $N\times N$ diagonal matrix $\Lambdabf$, we have
\beq[HPG]
\Hbf\Phibf_t\Gbf^H = \Hbf\Lambdabf\Phibf_t\Lambdabf^{-1}\Gbf^H = \tilde{\Hbf}\Phibf_t\tilde{\Gbf}^H \; ,
\eeq
where $\tilde{\Hbf}=\Hbf\Lambdabf$ and $\tilde{\Gbf}=\Gbf(\Lambdabf^*)^{-1}$. Thus there is a scaling ambiguity between each pair of the $N$ columns $\{\hbf_k,\gbf_k\}$ of $\Hbf$ and $\Gbf$ that cannot be resolved using data obtained as in~\eqref{sumimo}. Most methods for beamforming, precoding or RIS reflection optimization do not require this ambiguity to be resolved, although as briefly discussed later, with certain additional information the individual channel components can be identified. For this reason, channel estimation in the context of RIS-aided communication systems focuses primarily on determination of the $MK\times (N+1)$ {\em composite} or {\em cascaded} channel $\Hbf_c$, defined using properties of the Khatri-Rao product:
\beq[compchan]
{\rm vec}\left( \Hbf_d + \Hbf \Phibf_t \Gbf^H \right) = \left[ \hbf_d \; \; \Gbf^* \diamond \Hbf \right] \left[ \begin{array}{c} 1 \\ \phibf_t \end{array}
\right] \equiv \Hbf_c \tilde{\phibf}_t \; ,
\eeq
with $\hbf_d={\rm vec}(\Hbf_d)$. Eq.~\eqref{compchan} together with further use of the Kronecker product allows us to rewrite~\eqref{sumimo} in a compact form:
\begin{subequations}
\label{krmimo}
\begin{align}
\ybf_t &= \sqrt{P} \left( \xbf_t^T \otimes \Ibf_M \right) \Hbf_c \tilde{\phibf}_t + \nbf_t \label{krmimo1} \\
&= \sqrt{P} \left[ \tilde{\phibf}_t^T \otimes \xbf_t^T \otimes \Ibf_M \right] \hbf_c  + \nbf_t \\
&\equiv \sqrt{P}\Zbf_t \hbf_c + \nbf_t \; ,
\end{align}
\end{subequations}
where $\hbf_c = {\rm vec}(\Hbf_c)$ and the $M \times MK(N+1)$ matrix $\Zbf_t$ is implicitly defined.

The composite channel $\hbf_c$ is clearly underdetermined in Eq.~\eqref{krmimo}, and thus multiple pilot symbols must be transmitted in order for it to be uniquely estimated. Combining the data from $T$ such pilots together, we have 
\beq[fulltrain]
\mathbf{\mathfrak{y}} = \left[ \begin{array}{c} \ybf_1 \\ \vdots \\ \ybf_T \end{array} \right] = \sqrt{P} \left[ \begin{array}{c} \Zbf_1 \\ \vdots \\ \Zbf_T \end{array} \right] \hbf_c + \mathbf{\mathfrak{n}} \equiv \sqrt{P} \Zbf \hbf_c + \mathbf{\mathfrak{n}} \; .
\eeq
Provided that $T \ge K(N+1)$ and $\Zbf$ is full rank, there are two common ways to estimate $\hbf_c$, as discussed below.

\subsubsection{Least Squares} \label{sec:ls}
The simplest approach for estimating $\hbf_c$ is to use the standard deterministic least-squares (LS) method, 
\beq[lsest]
\hat{\hbf}_{c,LS} = \arg\min_{\hbf_c} \; \| \mathbf{\mathfrak{y}} - \sqrt{P} \Zbf \hbf_c \|^2 \; ,
\eeq
whose solution is given by
\beq[lssol]
\hat{\hbf}_{c,LS} = \frac{1}{\sqrt{P}} \Zbf^\dagger \mathbf{\mathfrak{y}} \; ,
\eeq
where $\Zbf^\dagger = \left(\Zbf^H\Zbf\right)^{-1} \Zbf^H$. Assuming again that $\nbf_t \sim \Cc\Nc(\zerobf,\sigma^2 \Ibf_M)$ and that the noise is temporally uncorrelated, the LS channel estimate is unbiased and equivalent to the maximum likelihood (ML) estimate, and its covariance matrix corresponds to the CRB:
\begin{subequations}
\label{lscov}
\begin{align}
\Rbf_{\hat{\hbf}_{c,LS}} &= \Ec\left\{ \left(\hat{\hbf}_{c,LS} - \hbf_c\right) \left(\hat{\hbf}_{c,LS} - \hbf_c\right)^H \right\} \\
&= \frac{1}{P} \Ec\left\{ \Zbf^\dagger \mathbf{\mathfrak{n}} \mathbf{\mathfrak{n}}^H (\Zbf^\dagger)^H \right\} = \frac{\sigma^2}{P} \left(\Zbf^H\Zbf\right)^{-1} \; .
\end{align}
\end{subequations}

Ideally, $\xbf_t$ and $\phibf_t$ should be designed to optimize the CSI estimation performance. While such an optimization is generally intractable, a good choice can be found \cite{LS-mvue} by noting that for any positive definite matrix $\Bbf$, we have 
\beq[bii] 
\left[ \Bbf^{-1} \right]_{ii} \ge \frac{1}{\Bbf_{ii}} \; , 
\eeq
with equality for all $i$ only if $\Bbf$ is diagonal. Thus, a good choice for $\Zbf$ would make~\eqref{lscov} diagonal. Such a choice may not be optimal in general, but a diagonal covariance matrix also greatly simplifies the computation of $\hat{\hbf}_{c,LS}$ in~\eqref{lssol}.

The most common training approach that meets the above design goals breaks the training interval $T$ into $T/K$ subblocks of length $K$, where $T/K$ is assumed to be an integer. For each subblocks, $b=1,\cdots,T/K$, $\phibf_t=\bar{\phibf}_b$ is held constant, while the pilots $\xbf_t$ are chosen as an orthonormal sequence that repeats itself for each subblock. For example, the subblock sequence for the UE is $\Xbf=\left[\xbf_1 \; \cdots \; \xbf_K\right]$, where $\Xbf\Xbf^H=K \Ibf_K$, which is then repeated $T/K$ times: 
\begin{subequations}
\label{blockpilots}
\begin{align}
    \text{$\xbf_t$ pilots} &= [ \underbrace{\Xbf \; \Xbf \; \cdots \; \Xbf}_{\text{repeated $T/K$ times}} ] \\[8pt]
    \text{$\phibf_t$ pilots} &= [ \underbrace{\bar{\phibf}_1 \; \cdots \; \bar{\phibf}_1}_{\text{repeated $K$ times}} \; \cdots \;
    \underbrace{\bar{\phibf}_{\frac{T}{K}} \; \cdots \; \bar{\phibf}_{\frac{T}{K}}}_{\text{repeated $K$ times}} ]
\end{align}
\end{subequations}
Using this approach, we have
\begin{subequations}
\begin{align}
\Zbf^H\Zbf &= \sum_{t=1}^T \left[ \tilde{\phibf}_t^*\tilde{\phibf}_t^T \otimes \xbf_t^*\xbf_t^T \otimes \Ibf_M \right] \\
&= \left( \sum_{t=1}^T \left[ \tilde{\phibf}_t^*\tilde{\phibf}_t^T \otimes \xbf_t^*\xbf_t^T \right] \right) \otimes \Ibf_M \\
&= \left( \sum_{b=1}^{T/K} \tilde{\bar{\phibf}}_b^*\tilde{\bar{\phibf}}_b^T \right) \otimes (\Xbf \Xbf^H)^* \otimes \Ibf_M \\
&= K \left( \Psibf^H \Psibf \right)^* \otimes \Ibf_{MK} \; ,
\end{align}
\end{subequations}
where $\tilde{\bar{\phibf}}_b^H=[ 1 \; \; \bar{\phibf}_b^H ]$ and
\beq[psidef]
\Psibf = \left[ \begin{array}{cc} 1 & \bar{\phibf}_1^H \\[4pt] \vdots & \vdots \\[4pt] 1 & \bar{\phibf}_{T/K}^H \end{array} \right] \; .
\eeq
To achieve a diagonal $\Psibf^H\Psibf$, the columns of the $\frac{T}{K} \times (N+1)$ matrix $\Psibf$ must be orthogonal, with $\frac{T}{K} \ge N+1$. If $\Psibf^H\Psibf$ can be made proportional to an identity matrix, then $\Zbf^H\Zbf$ is also a scaled identity matrix.

For the above training protocol, the general solution in~\eqref{lssol} is implemented by taking data from the $b$-th pilot subblock,
\beq[pilotsub]
\Ybf_b = \sqrt{P} \left( \Hbf_d + \Hbf \bar{\Phibf}_b \Gbf^H \right) \Xbf + \Nbf_b \; , 
\eeq
and multiplying on the right by $\Xbf^H/(K\sqrt{P})$ to obtain
\beq[eq115]
\mathbf{\mathfrak{y}}_b \equiv \frac{1}{K\sqrt{P}} \text{vec} \left(\Ybf_b \Xbf^H\right) = \Hbf_c \tilde{\bar{\phibf}}_b + \bar{\nbf}_b \; .
\eeq
where $\bar{\Phibf}_b={\rm diag}(\bar{\phibf}_b)$ and $\bar{\nbf}_b=\text{vec} \left(\Nbf_b \Xbf^H\right)/(K\sqrt{P})$. The result $\mathbf{\mathfrak{y}}_b$ from each of the $T/K$ subblocks then forms a column of the following combined equation:
\beq[HcPN]
\mathbf{\mathfrak{Y}}_c = \Hbf_c \left[ \tilde{\bar{\phibf}}_1 \; \cdots \; \tilde{\bar{\phibf}}_{\frac{T}{K}} \right] + \bar{\Nbf} = \Hbf_c \Psibf^H + \bar{\Nbf} \; ,
\eeq
where $\bar{\Nbf}=[\bar{\nbf}_1 \; \cdots \; \bar{\nbf}_{T/K} ]$, from which an estimate of the composite channel is obtained by multiplying $\mathbf{\mathfrak{Y}}_c$ by $\Psibf$ on the right, assuming $\Psibf^H\Psibf \propto \Ibf_{N+1}$. 

Several methods have been proposed to choose the RIS training sequence to satisfy $\Psibf^H\Psibf \propto \Ibf_{N+1}$:
\begin{itemize}
\item When the direct path is absent (the first column of $\Psibf$ is removed), a simple approach is to set $\frac{T}{K}=N$ and ``turn on'' one RIS element at a time for each $K$-sample pilot subblock, with all other elements ``turned off''\footnote{``Turning off'' an RIS element assumes it becomes a perfect absorber of RF energy, which in practice is not possible. Thus, such elements will still reflect a small amount of energy and thus degrade the orthogonality assumption.} \cite{2020Nadeem,KimL21}. This results in $\Psibf^H\Psibf = {\rm diag}\{\beta_1^2, \cdots, \beta_N^2\}$. If each (identical) RIS element when active is tuned to the same phase, it is reasonable to assume that $\beta_i=\beta$, which results in $\Zbf^H\Zbf = \beta^2 K \Ibf_{MKN}$ and an estimate variance of $\sigma^2/(\beta^2 P K)$ for each element of $\hbf_c$. 
\item Better performance is achieved by activating all RIS elements over the entire training interval, in order to benefit from the RIS array gain. One approach for doing so assigns the RIS phase shifts such that the $N+1$ columns of $\Psibf$ equal the columns of the $\frac{T}{K}\times \frac{T}{K}$ matrix that defines the $\frac{T}{K}$-point Discrete Fourier Transform (DFT) \cite{LS-mvue,2020Nadeem}:
\beq[dftphase]
\left[ \Psibf \right]_{mn} = e^{j2\pi (m-1)(n-1)/(T/K)} 
\eeq
for $m=1,\cdots,\frac{T}{K}$ and $n=1,\cdots,N+1$. If the RIS gains are assumed to be phase-independent and satisfy $\beta_i=\beta$, then this leads to $\Psibf^H\Psibf = \frac{T\beta^2}{K} \I_{N+1}$ and the variance of the channel coefficient estimates is $\sigma^2/(\beta^2 P T)$, a factor of $T/K \ge N+1$ smaller than in the first approach. In addition to the need for phase-independent RIS element gains, which is difficult to achieve in practice, the RIS phase shifts should be tunable with at least $\log_2(T/K)$ bits of resolution, which may be problematic for large $N$.
\item An alternative that achieves the same performance is to choose the columns of $\Psibf$ from among the columns of a $T/K$-dimensional Hadamard matrix, whose entries are constrained to be $\pm 1$ \cite{2020changsheng,Bjornson21}. This achieves orthogonality for $\Psibf$, and has the advantage of requiring only two phase states for each RIS element (one bit of resolution). In addition, a diagonal $\Psibf^H\Psibf$ only requires that the RIS gains be equal at these two phase values. In this approach, $T/K$ must be a multiple of 4 for the Hadamard matrix to exist, but this is not a significant issue for large $N$.
\end{itemize}

\subsubsection{Linear Minimum Mean Squared Error}

The LS approach assumes a deterministic channel with no prior information. On the other hand, the minimum mean-squared error (MMSE) estimator assumes a stochastic model for $\{\Hbf_d,\Gbf,\Hbf\}$, usually in terms of correlated Rayleigh fading with prior information of the second-order statistics. However, the composite channel is composed of products of the Gaussian elements in $\Hbf$ and $\Gbf$, which makes the MMSE estimate $\Ec\{\hbf_c | \mathbf{\mathfrak{y}} \}$ difficult to compute, although message-passing algorithms have been proposed for this problem \cite{2021ZhenQing,HeY20,MirzaA21}. Instead, the linear MMSE, or LMMSE, estimate given by $\hat{\hbf}_{c,LM} = \Wbf \mathbf{\mathfrak{y}}$ can be found by solving \cite{KunduM21,SumanKH21}
\beq[mmse]
\Wbf = \arg\min_{\tilde{\Wbf}} \Ec \left\{ \| \tilde{\Wbf}\mathbf{\mathfrak{y}} - \hbf_c \|^2 \right\} \; .
\eeq
Assuming spatially and temporally white Gaussian noise uncorrelated with $\hbf_c$, the LMMSE estimate is given by
\beq[mmsesol]
\hat{\hbf}_{c,LM} = \sqrt{P} \Rbf_{\hbf_c} \Zbf^H \left( P \Zbf\Rbf_{\hbf_c}\Zbf^H + \sigma^2 \Ibf_{MT} \right)^{-1} \mathbf{\mathfrak{y}} \; ,
\eeq
where $\Rbf_{\hbf_c}=\Ec\{\hbf_c \hbf_c^H\}$ and we have assumed $\Ec\{\hbf_c\}=\zerobf$. 

Using orthogonal pilot and RIS reflection sequences like those discussed above also simplifies computation of the LMMSE estimate. For example, let $\Ibf=\Ibf_{MK(N+1)}$ and assume the Hadamard reflection pattern so that $\Zbf^H\Zbf=T \Ibf$. Then the LMMSE estimate simplifies to
\beq[mmsesol2]
\hat{\hbf}_{c,LM} = \frac{1}{\sqrt{P}T} \Rbf_{\hbf_c} \left( \Rbf_{\hbf_c} + \frac{\sigma^2}{PT} \Ibf \right)^{-1} \Zbf^H \mathbf{\mathfrak{y}} \; .
\eeq
The matrices in~\eqref{mmsesol2} involving $\Rbf_{\hbf_c}$ are data independent, and can be computed and stored offline since $\Rbf_{\hbf_c}$ changes relatively slowly. The resulting error covariance is given by
\beq[mmserr]
\Rbf_{e,LM}=\Rbf_{\hbf_c} - \Rbf_{\hbf_c} \left( \Rbf_{\hbf_c} + \frac{\sigma^2}{PT} \Ibf \right)^{-1} \Rbf_{\hbf_c} \; .
\eeq
For the above training protocol, $\Rbf_{e,LM}$ converges to $\R_{e,LS}=(\sigma^2/(PT)) \Ibf$ for high SNR ({\em i.e.,} $\sigma^2/P \rightarrow 0$) or long training intervals ($T \rightarrow \infty$).

A bigger issue than the computational complexity of~\eqref{mmsesol2} is how to determine the composite channel covariance $\Rbf_{\hbf_c}$. In theory, the covariance could be estimated using simulations involving detailed propagation models of the environment, or by taking sample statistics of channel estimates obtained over a long period of time. However, the size of $\hbf_c$ means that such procedures would require a large amount of data. Instead, a more reasonable approach is to determine $\Rbf_{\hbf_c}$ based on covariance information about its constituent parts. For MIMO channels, it is commonly assumed that the multipath scattering at the source is uncorrelated with the scattering at the destination, which leads to the following descriptions:
\begin{subequations}
\begin{align}
    \Hbf &= \Rbf_{HB}^{\half} \tilde{\Hbf} \Rbf_{HR}^{\frac{H}{2}} \\
    \Gbf &= \Rbf_{GU}^{\half} \tilde{\Gbf} \Rbf_{GR}^{\frac{H}{2}} \\
    \Hbf_d &= \Rbf_{H_dB}^{\half} \tilde{\Hbf}_d \Rbf_{H_dU}^{\frac{H}{2}} \; ,
\end{align}
\end{subequations}
where the subscripts $\{B,R,U\}$ respectively correspond to BS, RIS, and UE, and indicate which side of the link the correlation matrix is associated with ({\em e.g.,} $\Rbf_{HB}$ is the correlation matrix for the BS-side of the channel $\Hbf$). The matrices $\{\tilde{\Hbf},\tilde{\Gbf},\tilde{\Hbf}_d\}$ are of the same dimensions as $\{\Hbf,\Gbf,\Hbf_d\}$ respectively, and are composed of uncorrelated $\Cc\Nc(0,1)$ elements. Under this model, it can be shown that the composite channel covariance matrix has the following form:
\beq[compcov]
\Rbf_{\hbf_c} = \left[ \begin{array}{cc} \Rbf_{H_dU}^T \otimes \Rbf_{H_dB} & \zerobf^T \\ \zerobf & \Rbf_R \otimes \Rbf_{GU}^T \otimes \Rbf_{HB}
\end{array} \right] \; ,
\eeq
where we define $\R_R= \Rbf_{GR}\odot\Rbf_{HR}^T$. 

Estimating the correlation matrices $\Rbf_{H_dU}, \Rbf_{H_dB}, \Rbf_{GU}$ and $\Rbf_{HB}$ is relatively straightforward since the BS and UEs have active transceivers that can collect and process data. However, determining the RIS-side correlation matrices $\Rbf_{GR}$ and $\Rbf_{HR}$ is problematic since the RIS is typically passive. Various assumptions can be made to further simplify $\Rbf_{\hbf_c}$. With uncorrelated scattering at the RIS, $\R_{GR}$ and $\R_{HR}$ can be taken as identity matrices, and $\R_{\hbf_c}$ is block diagonal with identical block entries except for the block associated with $\Hbf_d$. This greatly simplifies computation of~\eqref{mmsesol2}. If we go a step further and assume all channels exhibit uncorrelated Rayleigh fading, then the LMMSE estimate and error simplify to
\begin{subequations}
\begin{align}
\hat{\hbf}_{c,LM} = \frac{1}{\sqrt{P}T} &\left[ \begin{array}{cc} \nu_{H_d} \I_{MK} & \zerobf \\ \zerobf & \nu_{GH} \I_{MNK} \end{array} \right] \Zbf^H \mathbf{\mathfrak{y}} \\[5pt]
\Rbf_{e,LM} = \frac{\sigma^2}{PT} &\left[ \begin{array}{cc} \nu_{H_d} \I_{MK} & \zerobf \\ \zerobf & \nu_{GH} \I_{MNK} \end{array} \right] \; ,
\end{align}
\end{subequations}
where 
\begin{subequations}
\begin{align}
    \nu_{H_d} &= \frac{PT \sigma^2_{H_d}}{PT \sigma^2_{H_d}+\sigma^2} < 1 \\[5pt]
    \nu_{GH} &= \frac{PT \sigma^2_{G}\sigma^2_H}{PT \sigma^2_{G}\sigma^2_H+\sigma^2} < 1 \; ,
\end{align}
\end{subequations}
and $\sigma^2_{H_d}, \sigma^2_{H}, \sigma^2_{G}$ represent the variances of the channels $\Hbf_d, \H, \G$, respectively. Since $\nu_{H_d}$ and $\nu_{GH}$ are less than one, the LMMSE estimates have a smaller error than for LS, which is due to the exploitation of the prior statistical information. However, assumptions of uncorrelated fading are hard to justify in RIS-aided wireless systems, which are typically motivated by propagation environments with sparse propagation paths and frequent blockages. In these environments, the BS and RIS installations are envisioned to be in elevated positions away from nearby RF scatterers. This leads to low-rank channel correlation matrices and consideration of geometric models, as discussed in Section~\ref{sec:geometric}. 

\subsection{Wideband Single User MIMO}\label{sec:unofdm}

In wideband scenarios where the channel is frequency selective, we assume the UE transmits an OFDM signal composed of $N_c$ subcarriers from each of its $k$ antennas. The symbols are given by the rows of the $K\times N_c$ matrix $\Xbf_t^F=\left[ \xbf^F_{t,1} \; \; \cdots \; \; \xbf^F_{t,N_c} \right]$ in the frequency domain, where here $t$ is the OFDM symbol index. Prior to transmission, the data $\Xbf_t^F$ is first converted to the time domain using the $N_c\times N_c$ matrix $\Fbf^H$ that denotes the $N_c$-point inverse DFT: $\Xbf_t = \Xbf_t^F \F^H$, and then is appended with a cyclic prefix of length $L_{cp}$ that is longer than the maximum delay spread of the channel, $L$. At the BS, the cyclic prefix is removed, and the data are converted back to the frequency domain through multiplication by the DFT matrix $\F$. This generates a model essentially identical to~\eqref{sumimo} for each subcarrier $n$:
\beq[fdmimo]
\ybf^F_{t,n} = \sqrt{P} \left( \Hbf^F_{d,n} + \Hbf^F_n \Phibf_{t,n} \Gbf^{FH}_n \right) \xbf^F_{t,n} + \nbf^F_{t,n} \; , 
\eeq
where $\{\H^F_{d,n}, \H^F_n, \G^F_n\}$ represent the DFT at subcarrier $n$ for the UE-BS, RIS-BS, and UE-RIS channel impulse responses, respectively. Thus, one can employ the same estimation methods discussed above on a per-subcarrier basis, although to exploit the channel correlation in frequency and reduce the training overhead, pilot data is normally transmitted only on a subset of the subcarriers, and interpolation used to construct channel estimates for others \cite{2019beixiong}. An alternative approach proposed in \cite{ZhengYZ21} is to use shorter OFDM symbols during the training period.

Note that most prior work on RIS channel estimation with OFDM signals has assumed that the RIS reflection properties are frequency independent, {\em i.e.,} $\Phibf_{t,n}=\Phibf_t$, but this is generally true only for relatively narrow bandwidths \cite{CaiLLL20,YangLLLL21}. If one sets $\phibf_{t,n}$ to have desirable properties ({\em e.g.,} $\Zbf$ with orthogonal columns) at a particular subcarrier $n$, then in general those properties will not be inherited at other subcarriers. This issue motivates the design of RIS circuit architectures that have invariant properties across wider frequency bands.

An alternative to estimating the channels in the frequency domain and using interpolation is to directly estimate the channel impulse response. In the time domain, we represent the data received for sample $s$ of OFDM symbol $t$ as
\beq[tdmimo]
\ybf_{t,s} = \sqrt{P} \sum_{k=0}^{L-1} \Bigl( \Hbf_d(k) + \Hbf(k) \Phibf_{t,s-k} \Gbf^{H}(k) \Bigr) \xbf_{t,s-k} + \nbf_{t,s} \; , 
\eeq
where $\{\Hbf_d(k), \Hbf(k), \Gbf(k)\}_{k=0}^{L-1}$ represent the channel impulse responses and $L$ is the maximum number of taps. Defining $\hbf_d(k)=\text{vec}(\H_d(k))$ and $\hbf_c(k)=\text{vec}\left(\left[\hbf_d(k) \; \; \Gbf^*(k)\diamond \Hbf(k)\right] \right)$, after removal of the cyclic prefix we can write
\begin{subequations}
\begin{align}
    \ybf_{t,s} & = \sqrt{P} \sum_{k=0}^{L-1} \left[ \tilde{\phibf}_{t,s-k}^T \otimes \xbf_{t,s-k}^T \otimes \Ibf_M \right] \hbf_c(k)  + \nbf_{t,s} \\
    &= \sqrt{P} \sum_{k=0}^{L-1} \Z_{t,s-k} \hbf_c(k) + \nbf_{t,s} \\
    &= \sqrt{P} \Bigl[ \Z_{t,s} \; \, \Z_{t,s-1} \; \; \cdots \; \; \Z_{t,s-L+1} \Bigr] \mathbf{\mathfrak{h}}_c + \nbf_{t,s} \\
    \mathbf{\mathfrak{y}}_t &= \left[ \begin{array}{c} \ybf_{t,1} \\ \vdots \\ \ybf_{t,N_c} \end{array} \right] = \sqrt{P} \mathbf{\mathfrak{Z}}_t \mathbf{\mathfrak{h}}_c + \mathbf{\mathfrak{n}}_t \; ,
\end{align}
\end{subequations}
where $\mathbf{\mathfrak{h}}_c=\left[ \hbf^T_c(0) \; \; \cdots \; \; \hbf^T_c(L-1) \right]^T$ is the $LMK(N+1) \times 1$ vector containing all unknown channel coefficients, and $\mathbf{\mathfrak{Z}}_t$ is an $MN_c \times LMK(N+1)$ block-circulant matrix with first block row $\left[ \Z_{t,1} \; \, \Z_{t,N_c} \; \; \cdots \; \; \Z_{t,N_c-L+2} \right]$. Finally, assuming the channel is stationary over $T_o$ total OFDM symbols, we have
\beq[totofdm]
\mathbf{\mathfrak{y}} = \sqrt{P} \left[ \begin{array}{c} \mathbf{\mathfrak{Z}}_1 \\ \vdots \\ \mathbf{\mathfrak{Z}}_{T_o} \end{array} \right] \mathbf{\mathfrak{h}}_c + \mathbf{\mathfrak{n}} = \sqrt{P} \mathbf{\mathfrak{Z}} \mathbf{\mathfrak{h}}_c + \mathbf{\mathfrak{n}} \; .
\eeq
The time-domain approach assumes only pilot data is transmitted first, followed by payload data. The total number of pilot symbols required is $T=T_oN_c\ge KL(N+1)$. While more OFDM symbols are likely required for the frequency domain method to obtain the same channel estimation accuracy, this is offset by the fact that data and pilots can be transmitted together.

\subsection{Single Antenna Scenarios}

\subsubsection{Single Antenna UE}\label{sec:saue1}

The single-antenna UE case is often considered in the literature, since it simplifies the notation and reduces the algorithm complexity, but there is fundamentally little difference with the general multi-antenna UE case described above. The channel $\G$ becomes a $1\times N$ row vector that we denote by $\gbf^T$, while the direct channel $\Hbf_d$ is simply an $M \times 1$ vector $\hbf_d$. The pilot data received at the BS is given by
\beq[SAUE]
\ybf_t = \sqrt{P} \left(\hbf_d + \Hbf \, \text{diag}({\gbf}^*) \phibf_t\right) x_t + \nbf_t \; ,
\eeq
where the composite channel is now $\Hbf_c = \gbf^H \diamond \Hbf = \Hbf \, \text{diag}({\gbf^*})$. The training overhead in this case is reduced to $N+1$ samples. 

\subsubsection{Single Antenna BS and UE}
When both the BS and UE have only a single antenna, we denote the RIS-BS channel as the $N\times 1$ row vector $\hbf^T$, and write the BS output and composite channel as
\begin{subequations}
\begin{align}
y_t &= \sqrt{P} \hbf_c^T \tilde{\phibf}_t x_t + n_t \\[5pt]
\hbf_c^T &= \left[ h_d \; \; \, \gbf^H \odot \hbf^T \right] = \left[ h_d \; \; \, \bar{\hbf}_c^T \right] \; ,
\end{align}
\end{subequations}
where only $\bar{\hbf}_c$ is identifiable. 

\subsection{Multiple User Scenarios}

The models and approaches discussed above are easily generalized to the multiple UE case. Assuming UE $u$ has $K_u$ antennas for $u=1,\cdots,U$, then the model in~\eqref{sumimo} holds if we simply set $K=\sum_u K_u$ and all UE antennas transmit orthogonal pilot sequences. Some prior work has proposed that the users take turns transmitting pilots, in which case there is no change to the algorithms described above, but this only makes sense if one exploits the fact that each user's composite channel shares a common RIS-BS component $\H$. This idea will be explored further in the next subsection. For multicarrier signals, a scheme is required to allocate the pilot subcarriers to the UEs, but otherwise the channel estimation is the same. One implication for the LMMSE approach is that, assuming the channels for different UEs are uncorrelated, the matrices $\R_{GU}$ and $\R_{H_cU}$ will be block-diagonal.

\subsection{Reducing the Complexity and Training Overhead}\label{sec:reduce}

As noted already above, one of the key hurdles to overcome in CSI estimation for RIS-aided systems is the large required training overhead. Consequently, recent work has focused on a variety of methods to reduce this overhead, some of which is described below. The use of geometric channel models to reduce pilot overhead is reserved for Section~\ref{sec:geometric}.

\subsubsection{RIS Element Grouping}
A simple approach to reduce the number of pilots and estimation complexity is to assign identical phases to RIS elements with highly correlated channels \cite{2019beixiong,2020changsheng}. High channel correlation occurs when adjacent RIS elements are closely spaced; retaining the flexibility of arbitrary phase shifts for such elements provides minimal additional beamforming gains. Suppose groups of size $J$ are identifed, and assume for simplicity that $N'=N/J$ is an integer and no direct channel $\Hbf_d$ is present. Then we define $\phibf_t = \phibf'_t \otimes \mathbf{1}_J$, where $\phibf'_t$ is $N' \times 1$, and write
\beq[Jgroup]
\Hbf_c \phibf_t = \Hbf_c (\phibf'_t \otimes \mathbf{1}_J) = \Hbf_c (\I_{N'} \otimes \mathbf{1}_J) \phibf'_t = \Hbf'_c \phibf'_t \, ,
\eeq
where the effective composite channel $\Hbf'_c$ is now $MK \times N'$. Each column of $\Hbf'_c$ is thus a unit-coefficient linear combination of the columns of $\Hbf_c$ corresponding to a given group of RIS elements. The revised model is identical in form to the general case, and thus the methods described above can be implemented to estimate $\Hbf'_c$ with a reduction in the required training overhead by a factor of $J$. A generalization of this idea presented in \cite{2020changsheng} successively reduces the size of the groups over multiple blocks of pilot and payload data in order to eventually resolve the channels for all of the RIS elements. 

\subsubsection{Low-Rank Channel Covariance}
We see from the noise-free part of~\eqref{fulltrain}, $\mathbf{\mathfrak{y}}=\sqrt{P}\Zbf\hbf_c$, that in the general case, the $MT\times MK(N+1)$ data matrix $\Zbf$ should be full rank $MK(N+1)$, since otherwise components of $\hbf_c$ in the nullspace of $\Zbf$ could not be identified. Like the LS approach, this requires $T\ge K(N+1)$ training samples. However, if $\R_{\hbf_c}$ is rank deficient, then it would be enough for the column span of $\R_{\hbf_c}$ to lie within the column span of $\Z^T$. In particular, suppose $\R_{\hbf_c}$ is rank $r < MK(N+1)$, and thus can be factored as $\R_{\hbf_c}=\U\U^H$, where $\U$ has $r$ columns. Then in principle it would be sufficient to choose 
\beq[ZT]
\Z^T = \U \V 
\eeq
for some full rank $r\times MT$ matrix $\V$, and thus theoretically it would be sufficient that $T \ge r/M$. Unfortunately, due to constraints on the possible values for $\phibf_t$, finding a $\V$ that exactly satisfies~\eqref{ZT} is generally not possible if $T < K(N+1)$. It may however be possible to approximately solve~\eqref{ZT} for larger values of $T$ that are still much smaller than $K(N+1)$, provided that $r$ is not too large. In addition to reducing the training overhead, the low rank channel covariance can be exploited to significantly reduce the cost of computing the LMMSE solution in~\eqref{mmsesol}, since only an $r\times r$ inverse rather than an $MT\times MT$ inverse is required:
\beq[mmserank]
\hat{\hbf}_c = \frac{\sqrt{P}}{\sigma^2} \U \left[ \I_r-\W \left( \W +\frac{\sigma^2}{P} \I_r \right)^{-1} \right] \U^H\Z^H \mathbf{\mathfrak{y}} ,
\eeq
where $\W=\U^H\Z^H\Z\U$.

\subsubsection{Exploiting Common Channels}
The LS method in Section~\ref{sec:ls} ignores the Kronecker product structure of the composite channel, which can be exploited to reduce the training overhead. The key observation is that, in the uplink, the composite channel for each user shares the same RIS-BS channel $\H$ \cite{2020Beixiong}. To explain how this information can be exploited, assume without loss of generality a scenario with $K$ single-antenna users. The approach is divided into two steps \cite{2020Zhaorui,2021LiWei}. In the first, one of the users is selected and the composite channel for this user is estimated in the normal way, while the other users do not transmit. Then, in the second step, the other users transmit and the estimate of the RIS-BS channel obtained in the first step is exploited to reduce the training required for the remaining channels.

Assume the users are ordered such that the user corresponding to the first row of $\Gbf$, denoted by $\gbf_1^T$, is the one selected for the first step. The LS method is used to estimate the composite channel $\Hbf\,\text{diag}(\gbf_1)$ and the direct channel $\hbf_{d,1}$, which requires at least $T_1=N+1$ training samples. Recall that only the product $\Hbf\,\text{diag}(\gbf_1)$ is estimated and not the individual terms $\Hbf$ and $\gbf_1$. In fact, we can treat $\text{diag}(\gbf_1)$ as $\Lambdabf$ in~\eqref{HPG}, so step~1 provides us with an estimate of $\tilde{\Hbf}$, and we can set the first row of $\tilde{\Gbf}^H$ to $\mathbf{1}_K$. With the estimate $\hat{\tilde{\Hbf}}$, during step~2 the training data model is approximately given by 
\begin{subequations}
\label{sumimo2}
\begin{align}
\ybf_t &\simeq \sqrt{P} \left( \tilde{\Hbf}_d + \hat{\tilde{\Hbf}} \Phibf_t \tilde{\Gbf}^H \right) \xbf_t + \nbf_t \\
&\simeq \sqrt{P} \underbrace{\xbf_t^T \otimes \left[ \I_M \; \; \, \hat{\tilde{\H}}\Phibf_t \right]}_{M\times (K-1)(M+N)} 
\left[ \begin{array}{c} \tilde{\hbf}_d \\ \tilde{\gbf}^* \end{array} \right] + \nbf_t \\
&\simeq \sqrt{P} \tilde{\Z}_t \tilde{\hbf}_c + \nbf_t \; ,
\end{align}
\end{subequations}
where we drop the first column of $\Hbf_d$ to create $\tilde{\Hbf}_d$, and we drop the first row of ones in $\tilde{\Gbf}$, since UE 1 does not transmit. We also have defined $\tilde{\gbf}= \text{vec}(\tilde{\Gbf}^T)$ and $\tilde{\hbf}_c^T = \left[ \tilde{\hbf}_d^T \; \; \, \tilde{\gbf}^H\right]$. Stacking $T_2$ of these training vectors together, we get an equation analogous to~\eqref{fulltrain}, where in this case $\Z$ is $MT_2\times (M+N)(K-1)$. Assuming linearly independent pilots $\xbf_t$ and RIS reflection vectors $\phibf_t$ are chosen, we can solve for the remaining channel parameters using $\hat{\tilde{\hbf}}_c = \Z^\dagger\mathbf{\mathfrak{y}}/\sqrt{P}$ provided that $MT_2 \ge (K-1)(M+N)$, or equivalently, $T_2\ge (K-1)(\frac{N}{M}+1)$. Given the $N+1$ samples needed for step~1, the minimum required training time is thus 
\beq[mintrain]
T_{min}=(N+1) + \left(\frac{N}{M}+1\right)(K-1) \; ,
\eeq
which for large $M$ is significantly less than the value $K(N+1)$ required by the standard LS method.

\section{Estimation of Structured Channels}\label{sec:geometric}

The large training overhead required for unstructured channel estimation motivates the consideration of channel models that are described by fewer parameters. Such models are often used in millimeter wave or higher frequency bands, where multipath scattering is sparse and propagation is often dominated by strong specular components. In such cases, the channels can be described by a small number of propagation paths defined by path gains, angles of arrival (AoAs), and angles of departure (AoDs)\footnote{For very large RIS, where the BS or UEs are in the Fresnel region of the RIS, the channel parameterization must also include range or the 3-D coordinates of the various devices, and the large scale fading becomes antenna-dependent. However, here we focus on the more common far-field scenario.}. The resulting number of parameters is often more 1-2 orders of magnitude less than that required in the unstructured case, and the training overhead is correspondingly reduced.

Parametric channels are described by the array response or ``steering'' vectors associated with the angle of an incoming (AoA) or outgoing (AoD) signal. For example, the response of an $M_x$-element uniform linear array (ULA) to a signal arriving with azimuth angle $\theta_{az}$ is described by the Vandermonde vector
\beq[stvechor]
\abf_{x}(\omega_x) = [1 \; \, e^{j\omega_x} \; \, e^{j2\omega_x} \;\, \cdots \;\, e^{j(M_x-1)\omega_x}]^T \; ,
\eeq
where the spatial frequency $\omega_x$ is defined by $\omega_x=2\pi \Delta_x\sin(\theta_{az})$, and $\Delta_x$ is the distance in wavelengths between the antennas\footnote{Note that we assume a narrowband propagation model here where time delays can be represented by phase shifts. For large arrays, ignoring the frequency dependence of the model leads to the beam-squint effect \cite{MaSAH21}.}. For an $M_x \times M_y$ uniform rectangular array (URA) with antenna separations of $\Delta_x$ and $\Delta_y$ in the $x$ and $y$ directions, the array response vector can be written as
\beq[svazel]
\abf(\omegabf) = \abf_{x}(\omega_x) \otimes \abf_{y}(\omega_y) \; ,
\eeq
where the vertical array response component is similar to~\eqref{stvechor},
\beq[stvecver]
\abf_{y}(\omega_y) = [1 \; \, e^{j\omega_2} \; \, e^{j2\omega_2} \;\, \cdots \;\, e^{j(M'_y-1)\omega_2}]^T \; ,
\eeq
but defined by $\omega_y=2\pi \Delta_y\sin(\theta_{el})\cos(\theta_{az})$ with elevation AoA $\theta_{el}$. The vector $\omegabf=[\omega_x \; \, \omega_y]^T$ corresponds to a 2D spatial frequency. For either a ULA or URA, there is a one-to-one correspondence between the angles and spatial frequencies as long as $\{\Delta_x,\Delta_y\}$ are no more than one-half wavelength. This is important for applications involving localization, since the angles provide useful information for locating a signal source. However, from the viewpoint of channel estimation, it is enough to know $\omegabf$, and any ambiguities in determining the angles need not be resolved.

In this section we focus on estimation of structured or geometric channel models. To simplify the discussion, we assume that the direct UE-BS channel is absent. This assumption is typical for scenarios with low-rank near-specular propagation at high frequencies, where blockages are common. The extra steps and pilot data required to estimate $\H_d$ when it is present generate minimal additional overhead. We will further assume that the BS and the UEs (when they have multiple antennas) employ ULAs, so that their array response depends on a single angle/spatial frequency, and we assume that the RIS elements are arranged as a URA, so its spatial response depends on two spatial frequencies. Generalizations to arbitrary array geometries are straightforward.

\subsection{Parametric Estimation}

The structured channel estimators that we will consider assume parametric channel models of the following form, which we describe first for the RIS-BS channel:
\begin{subequations}
\label{Hparam}
\begin{align}
\Hbf &= \sum_{k=1}^{d_H} \gamma_{H,k} \abf_B(\omegabf_{BH,k}) \abf_U^H(\omegabf_{RH,k}) \\
&= \Abf_B(\omegabf_{BH}) \Gammabf_H \Abf_R^H (\omegabf_{RH}) \; ,
\end{align}
where the columns of
\begin{align}
\Abf_B(\omegabf_{BH}) &= \left[ \abf_B(\omega_{BH,1}) \;\; \cdots \;\; \abf_B(\omega_{BH,d_H}) \right] \\
\Abf_R(\omegabf_{RH}) &= \left[ \abf_R(\omegabf_{RH,1}) \;\; \cdots \;\; \abf_R(\omegabf_{RH,d_H}) \right] 
\end{align} 
\end{subequations}
respectively represent the steering vectors for the propagation paths with AoAs $\omegabf_{BH}=[\omega_{BH,1} \; \cdots \; \omega_{BH,d_H}]^T$ at the BS and AoDs $\omegabf_{RH}=[\omegabf_{RH,1}^T \; \cdots \; \omegabf_{RH,d_H}]^T$ from the RIS. The diagonal matrix $\Gammabf_H={\rm diag}\{ \gammabf_H\}={\rm diag}\{\gamma_{H,1} \; \, \cdots \; \, \gamma_{H,d_H}\}$ contains the complex path gains $\gammabf_H=[\gamma_{H,1} \; \, \cdots \; \, \gamma_{H,d_H}]^T$. The RIS AoDs for path $k$, denoted by $\omegabf_{RH,k}$, are written as vectors since the RIS spatial frequencies are two-dimensional:
\beq[azel]
\omegabf_{RH,k} = \left[ \begin{array}{c} \omega_{RH,k,x} \\[5pt] \omega_{RH,k,y} \end{array} \right] \; .
\eeq
Parametric models like \eqref{Hparam} are usually employed when the number of paths $d_H$ is smaller than the array dimensions $M$ and $N$, and thus the channel $\Hbf$ is low-rank.

Parametric CSI estimation involves finding the spatial frequencies of signals collected by an array. For example, suppose $n$ observations are available from an arbitrary $M'$-element array receiving signals from $d$ directions:
\beq[ASN]
\Ybf' = \Abf(\omegabf') \Sbf' + \Nbf' \; ,
\eeq
where $\Ybf'$ is $M'\times n$, $\Sbf'$ is $d\times n$, $\Nbf'$ is noise, $\Abf=[\abf(\omegabf'_1) \; \; \cdots \; \; \abf(\omegabf'_d)]$ is the $M'\times d$ array response matrix, and $M'>d$. The matrix $\S'$ is not typically assumed to be known. This is the classical model assumed for AoA estimation, and many methods have been developed to estimate $\omegabf'=[\omegabf'_1 \; \; \cdots \omegabf'_d]^T$ from $\Y'$. The simplest method is based on (matched filter) beamforming, which involves searching for $d$ peaks in the spectrum
\beq[bfspec]
p_B(\omegabf) = \abf^H(\omegabf)\Rbf_{Y'}\abf(\omegabf) \; ,
\eeq
where $\R_{Y'}$ is the sample covariance matrix
\beq[ryy]
\Rbf_{Y'} = \frac{1}{n} \Ybf' \Ybf^{'H} \; .
\eeq
Alternatively, one can employ higher resolution algorithms such as the MUSIC \cite{music} or ESPRIT \cite{esprit}, which require computation of the eigendecompositon of $\R_{Y'}$. The beamforming and MUSIC spectra are either one- or two-dimensional functions, depending on whether the steering vectors depend on one- or two-dimensional spatial frequencies. If $\N'$ is spatially and temporally white, the (deterministic) maximimum likelihood (ML) method \cite{StoicaS90} finds the AoA estimates from the $d$-dimensional (or $2d$-dimensional for azimuth/elevation angles) problem
\beq[dml]
\hat{\omegabf}'_{ML} =\arg\min_\omegabf \text{trace} \left( \P_{A}^\perp(\omegabf) \R_{Y'} \right) \; ,
\eeq
where $\P_{A}^\perp(\omegabf)=\I_{M'}-\A(\omegabf)\left[ \A^H(\omegabf)\A(\omegabf)\right]^{-1} \A^H(\omegabf)$. The corresponding ML estimate of $\Sbf'$ is given by $\hat{\S}'=\Abf^\dagger(\hat{\omegabf}'_{ML})\Y'$, where $(\cdot)^\dagger$ represents the pseudo-inverse, although estimates of $\hat{\omegabf}'$ from other algorithms can be substituted for $\omegabf'_{ML}$ to estimate $\S'$. MUSIC and ESPRIT require that $\text{rank}(\Sbf')=d$ and thus that $n \ge d$, but the beamforming and ML methods are theoretically viable for any value of $n$, including $n=1$. For a massive antenna array where $M'$ is large, all of the above methods provide asymptotically efficient AoA estimates \cite{VibergON95}, and thus a simple technique such as beamforming is preferred due to its low computational load and minimal assumptions. A key requirement for all of the above methods is that the value of $d$ be known or estimated from the data.

\subsection{Compressive Sensing}

Compressive sensing (CS) formulations of the geometric CSI estimation problem are also possible, using sparse representations from an overcomplete dictionary \cite{TangN10,Bilik11,GurbuzCM12,ShenLCW16}. For example, we can represent the array response vectors for the BS side as
\beq[BSover]
\Abf_{B}(\omegabf_{BH}) = \Abf_{BD} \Qbf_{BH} \; ,
\eeq
where $\Abf_{BD}$ is an $M\times N_{BD}$ matrix whose columns are BS array response vectors sampled on an $N_{BD}$ grid of frequencies corresponding to the possible BS AoAs, and $\Qbf_{BH}$ is an $N_{BD}\times d_H$ matrix whose $k$-th column has a single 1 in the position corresponding to $\omega_{BH,k}$, assuming it is one of the grid points. If $\omega_{BH,k}$ is not on the grid, then the model in~\eqref{BSover} is an approximation. While the error can be made small by making $N_{BD}$ large, increasing the coherence of the dictionary eventually leads to computational and numerical issues. More will be said on this topic below. The RIS also has a similar overcomplete representation:
\beq[RISover]
\Abf_{R}(\omegabf_{RH}) = \Abf_{RD} \Qbf_{RH} \; ,
\eeq
where $\Abf_{RD}$ is $N\times N_{RD}$ and $\Qbf_{RH}$ is $N_{RD}\times d_H$. Because the RIS AoDs are two dimensional, a two-dimensional grid is necessary to specify the AoD pairs, and hence $N_{RD}=O(N_{BD}^2)$. Substituting~\eqref{BSover} and~\eqref{RISover} into~\eqref{Hparam}, we can write
\begin{subequations}\label{Hover}
\begin{align}
    \Hbf &=\Abf_{BD} \Qbf_{BH}\Gammabf_H \Qbf_{RH}^T \Abf^H_{RD} \\
    \mathbf{\mathfrak{h}} &= \text{vec}\left(\Hbf\right) = \left(\Abf^*_{RD} \otimes \Abf_{BD}\right) \gammabf_{HD} \; ,
\end{align}
\end{subequations}
where $\gammabf_{HD}=\text{vec} (\Qbf_{BH}\Gammabf_H\Qbf_{RH}^T)$ is a $d_H$-sparse vector of length $N_{BD}N_{RD}$ whose $d_H$ non-zero elements are equal to $\gammabf_H$. The $MN \times N_{BD}N_{RD}$ matrix $\Abf^*_{RD} \otimes \Abf_{BD}$ can be thought of as an overcomplete dictionary for the vectorized channel $\mathbf{\mathfrak{h}}$. 

Consider an overcomplete representation of the data in~\eqref{ASN}:
\beq[overasn]
\Y' = \A_D \Q'\S' + \N' = \A_D \C' + \N' \; ,
\eeq
where the dictionary $\A_D$ is $M'\times N_D$ and the matrix $\Q'$ is $N_D \times d$, with one non-zero element per column equal to one. As before $\S'$ is $d\times n$. The matrix $\C'=\Q'\S'$ exhibits common row sparsity; only $d$ rows have non-zero elements, and these rows are equal in some order to the rows of $\S'$. If $\C'$ can be estimated, then the location of its non-zero rows would correspond to the AoAs, and the entries in those rows to the rows of $\S'$. The problem can be solved, for example, by means of a convex LASSO-type approach of the form
\beq[lasso]
\hat{\C}' = \arg\min_\C \; \frac{1}{2} \| \Y' - \A_D \C\|^2_2 + \lambda \|\C\|_{2,1} \; ,
\eeq
where $\C^T = \left[ \cbf_1 \; \; \cdots \; \; \cbf_{N_D} \right]$ and the regularization
\beq[21norm]
\|\C\|_{2,1} = \sum_{k=1}^{N_D} \|\cbf_k\|_2
\eeq
promotes row sparsity. As mentioned above, the accuracy of approaches like~\eqref{lasso} is limited by the resolution of the sampled grid, which cannot be made infinitely fine due to numerical and computational issues. Off-grid AoAs create a basis mismatch that leads to leakage of energy into adjacent rows of $\C'$. One approach to refine the AoA estimates obtained by~\eqref{lasso} is to apply a small angular rotation to each selected column of $\A_D$ that maximizes the correlation with the received data \cite{ZhouPRPS21}:
\beq[rot]
\omegabf'_k = \arg\max_\omegabf \left\| {\rm diag}\left(\abf(\omegabf)\right) \abf_{D,k} \Ybf^{'H} \right\|_2  
\eeq
for $k=1,\cdots, d,$ where $\abf_{D,k}$ is the $k$-th column of $\A_D \hat{\Q}'$. The final estimate of $\omegabf_k$ is the spatial frequency corresponding to $\abf_{D,k}$ plus the estimated rotation $\omegabf'_k$. A more fundamental approach to solving the basis mismatch problem for one-dimensional spatial frequencies is to recast the sparse recovery problem using the atomic norm \cite{YangX15,2021Jiguang}. 

\subsection{Single User MIMO Single Carrier}\label{sec:sumimosc}

In this section, we consider structured channel estimation for the case involving a single multi-antenna UE. As with the RIS-BS channel, we can define parametric and overcomplete representations of the RIS-UE channel as follows:
\begin{subequations}
\label{Gparam}
\begin{align}
    \Gbf &= \Abf_U(\omegabf_{UG}) \Gammabf_G \Abf_R^H (\omegabf_{RG}) \\
    &= \Abf_{UD} \Qbf_{UG}\Gammabf_G \Qbf_{RG}^T \Abf^H_{RD} \\[5pt]
    \mathbf{\mathfrak{g}} & = \text{vec}\left(\Gbf\right) = \left(\Abf^*_{RD} \otimes \Abf_{UD}\right) \gammabf_{GD} \; ,
\end{align}
\end{subequations}
where we assume $d_G$ paths with gains $\gammabf_G=[\gamma_{G,1} \; \, \cdots \; \, \gamma_{G,d_G}]^T$ and $\Gammabf_G={\rm diag}\{ \gammabf_G\}$. In this case, $\gammabf_{GD}=\text{vec} (\Qbf_{UG}\Gammabf_G\Qbf_{RG}^T)$ is a $d_G$-sparse vector of length $N_{UD}N_{RD}$ whose non-zero elements correspond to $\gammabf_G$.

Recall the general model in~\eqref{fulltrain}, where the BS data from the $T$ training samples is stacked together in a single vector:
\beq[fulltrain2]
\mathbf{\mathfrak{y}} = \sqrt{P} \Zbf \hbf_c + \mathbf{\mathfrak{n}} \; ,
\eeq
where 
\beq[Zdef]
\Zbf = \left[ \begin{array}{c} \phibf_1^T \otimes \xbf_1^T \otimes \I_M \\ \vdots \\ \phibf_T^T \otimes \xbf_T^T \otimes \I_M \end{array} \right]
\eeq
is $MT\times MKN$ and involves $\phibf_t$ instead of $\tilde{\phibf}_t$ since we are assuming no direct UE-BS channel is present, which also implies that $\hbf_c=\text{vec}(\Hbf_c)=\text{vec}\left(\G^* \diamond \H\right)$.
Using various properties of the Kronecker and Khatri-Rao products, the composite channel can be decomposed using either a parametric approach or via overcomplete dictionaries as follows:
\begin{subequations}
\label{Hcab}
\begin{align}
\Hbf_c & = \underbrace{\left(\A^*_U(\omegabf_{UG}) \otimes \Abf_B(\omegabf_{BH})\right)}_{MK\times d_Hd_G} \underbrace{\Gammabf_{GH}}_{d_Hd_G\times d_Hd_G} \nonumber \\
& \qquad \qquad \qquad \qquad \times \underbrace{\left(\Abf_R^T(\omegabf_{RG})  \diamond \Abf_R^H(\omegabf_{RH})\right)}_{d_Hd_G \times N} \\[5pt]
 & = \underbrace{\left(\A^*_{UD} \otimes \Abf_{BD} \right)}_{MK\times N_{UD}N_{BD}} \underbrace{\Qbf_{GH}}_{N_{UD}N_{BD} \times N_{RD}^2} \underbrace{\left( \Abf_{RD}^T \diamond \Abf_{RD}^H \right)}_{N_{RD}^2 \times N} \; ,
\end{align}
\end{subequations}
where
\begin{subequations}
\begin{align}
    \Gammabf_{GH} &=\Gammabf^*_G \otimes \Gammabf_H \\[5pt]
    \Q_{GH} & = \underbrace{\left( \Q_{UG}\Gammabf^*_G \Q_{RG}^T\right)}_{N_{UD}\times N_{RD}} \otimes \underbrace{\left(\Q_{BH}\Gammabf_H\Q_{RH}^T\right)}_{N_{BD}\times N_{RD}} \label{QGHdef}\\
    &= \underbrace{(\Q_{UG}\otimes\Q_{BH})}_{N_{UD}N_{BD} \times d_Hd_G} \Gammabf_{GH} \underbrace{(\Q_{RG}\otimes\Q_{RH})^T}_{d_Hd_G \times N_{RD}^2} \; .
\end{align}
\end{subequations}

As in the non-parametric case, not all of the parameters or decompositions shown in~\eqref{Hcab} are identifiable. To see this, let
\beq[ldiag]
\Lambdabf={\rm diag}\left(\abf_R(\omegabf')\right)
\eeq
be a diagonal matrix formed from an arbitrary RIS array response vector for $2 \times 1$ DOA $\omegabf'$. Then,
\begin{align}
\Abf_R^T(\omegabf_{RG})  \diamond \Abf_R^H(\omegabf_{RH}) &= \Abf_R^T(\omegabf_{RG})\Lambdabf  \diamond \Abf_R^H(\omegabf_{RH})\Lambdabf^* \nonumber \\
&= \Abf_R^T(\omegabf'_{RG})  \diamond \Abf_R^H(\omegabf'_{RH}) \label{lambi}
\end{align}
where the elements of $\omegabf'_{RH}$ and $\omegabf'_{RG}$ are defined as $\omegabf'_{RH,k}=\omegabf_{RH,k}+\omegabf'$ and $\omegabf'_{RG,k}=\omegabf_{RG,k}+\omegabf'$. Thus, if the RIS AoA and AoD spatial frequencies are shifted by the same amount, there is no change to the composite channel response. In addition, since the channel gains for $\Hbf$ and $\Gbf$ always appear together as $\Gammabf_G^*\Gammabf_H$, there is a scaling ambiguity; in particular $\gammabf_H$ and $\gammabf_G$ yield the same composite channel as $\alpha \gammabf_H$ and $\gammabf_G / \alpha^*$. To obtain an identifiable parameterization $\etabf$ for the composite channel in~\eqref{Hcab}, one could for example set $\omegabf_{RH,1}=[0 \; \, 0]^T$ so that $\abf_R(\omegabf_{RH,1})$ is a vector of ones, and set $\gammabf_{H,1}=1$. With these assumptions, the set of unique parameters $\etabf$ that describe the composite channel matrix are given in Table~\ref{tab:param}. The total number of parameters is thus ${\rm dim}(\etabf)=5d_H+5d_G-4$, which for typical values of $d_H$ and $d_G$ is much smaller than the number of parameters $2MKN$ that must be estimated for a non-parametric $\Hbf_c$, which could be in the thousands.

\begin{table}
    \centering
    \begin{tabular}{c|c|c}
    {\sc Variable} & {\sc Parameters} & {\sc Description} \\ \hline
    $\omegabf_{BH}$ & $d_H$ & AoA frequencies at BS \\
    $\omegabf_{RH}$ & $2(d_H-1)$ & AoD frequencies at RIS \\
    $\gammabf_H$ & $2(d_H-1)$ & complex gains for BS-RIS channel \\
    $\omegabf_{RG}$ & $2d_G$ & AoA frequencies at RIS \\ 
    $\omegabf_{UG}$ & $d_G$ & AoD frequencies at UE \\
    $\gammabf_G$ & $2d_G$ & complex path gains for RIS-UE channel \\ \hline
    \end{tabular}
    \caption{Composite Channel Parameterization for Single User MIMO Single Carrier Case. The elements of these vectors form the elements of the parameter vector $\etabf$ for geometric channel models.}
    \label{tab:param}
    \vskip -0.5cm
\end{table}

\subsubsection{Channel Estimation for the General Case}
Using~\eqref{Hcab}, the composite channel is given by
\begin{subequations}
\label{Hcabvec}
\begin{align}
    \hbf_c &= \A(\omegabf) \gammabf_{GH} \label{haoagen} \\
    &= \A_D \qbf_{GH} \label{hcsgen} \; ,
\end{align}
\end{subequations}
where $\gammabf_{GH}=\gammabf_G^*\otimes\gammabf_H$, $\qbf_{GH}=\text{vec}(\Q_{GH})$, 
\begin{subequations}
\begin{align}
    \A(\omegabf) &= \Bigl(\Abf_R^T(\omegabf_{RG})  \diamond \Abf_R^H(\omegabf_{RH})\Bigr)^T \nonumber \\
    & \qquad \qquad \qquad \quad \diamond \left(\A^*_U(\omegabf_{UG}) \otimes \Abf_B(\omegabf_{BH})\right) \\[5pt]
    \A_D &= \left( \Abf_{RD}^T \diamond \Abf_{RD}^H \right)^T \otimes \left(\A^*_{UD} \otimes \Abf_{BD} \right) \label{ADdef}\; .
\end{align}
\end{subequations}
and $\omegabf^T=\left[ \omegabf_{BH}^T \; \; \omegabf_{RH}^T \; \; \omegabf_{RG}^T \; \; \omegabf_{UG}^T \right]$.
Similarly, there are two forms of the general data model in~\eqref{fulltrain2}:
\begin{subequations}
\label{aoacsgen}
\begin{align}
\mathbf{\mathfrak{y}} &= \sqrt{P} \Zbf \A(\omegabf) \gammabf_{GH} + \mathbf{\mathfrak{n}} \label{aoagen} \\
&= \sqrt{P} \Zbf \A_D \qbf_{GH} + \mathbf{\mathfrak{n}} \label{csgen} \; .
\end{align}
\end{subequations}

At first glance, Eq.~\eqref{aoagen} has the standard form assumed in AoA estimation problems, where $\A(\omegabf')=\Zbf\A(\omegabf)$, but there are some caveats. First, $\gammabf_{GH}$ is not arbitrary, but is instead a nonlinear function of $\gammabf_H$ and $\gammabf_G$. Second, note that the $k$-th column of $\A(\omegabf)$ can be expressed as
\beq[Aok]
\left[ \A(\omegabf) \right]_{:k} = \left[ \C^T(\omegabf_{RG},\omegabf_{RH}) \right]_{:k} \otimes \abf^*_U(\omega_{UG,\ell}) \otimes \abf_B(\omega_{BH,p}) ,
\eeq
where $\ell=\lfloor k/d_H \rfloor$, $p=\text{mod}_{d_H}(k)$, and
\beq[Cdef]
\C(\omegabf_{RG},\omegabf_{RH}) = \Abf_R^T(\omegabf_{RG})  \diamond \Abf_R^H(\omegabf_{RH}) \; .
\eeq
Each column of $\C^T(\omegabf_{RG},\omegabf_{RH})$ only depends on a pair of 2D angles, one each from $\omegabf_{RG}$ and $\omegabf_{RH}$ \cite{ZhouPRPS21}:
\begin{subequations}
\label{Cstruct}
\begin{align}
\left[ \C^T(\omegabf_{RG},\omegabf_{RH}) \right]_{:k} &= \left[ \left(\Abf_R^T(\omegabf_{RG})  \diamond \Abf_R^H(\omegabf_{RH})\right)_{k:}\right]^T \\
&= \left[ \abf^T_R(\omegabf_{RG,\ell}) \odot \abf^H_R(\omegabf_{RH,p}) \right]^T \\ 
&= \abf_R(\omegabf_{RG,\ell}-\omegabf_{RH,p}) \; .
\end{align}
\end{subequations}
Thus, from~\eqref{Aok}, we see that the $k$-th composite steering vector in~\eqref{aoagen} is parameterized by different pairs of entries from $\omegabf_{RG}$ and $\omegabf_{RH}$, and from $\omegabf_{BH}$ and $\omegabf_{UG}$:
\beq[Aok2]
\left[ \A(\omegabf) \right]_{:k} = \abf_R(\omegabf_{RG,\ell}-\omegabf_{RH,p}) \otimes \abf^*_U(\omega_{UG,\ell}) \otimes \abf_B(\omega_{BH,p}) \; .
\eeq

The fact that $\A(\omegabf)$ only depends on the differences between the elements of $\omegabf_{RG}$ and $\omegabf_{RH}$ is a direct consequence of the fact that they are not separately identifiable, as mentioned above. An AoA estimation algorithm that takes the special structure of $\gammabf_{GH}$ into account is difficult to formulate, and one-dimensional methods such as beamforming and MUSIC cannot exploit the inherent relationship between the columns of $\A(\omegabf)$. However, assuming $\Abf(\omegabf)$ is full rank\footnote{The matrix $\A(\omegabf)$ will generically be full rank as long as the BS, RIS, and UE arrays have unambiguous array manifolds ({\em e.g.}, elements spaced no more than $\lambda/2$ apart for a ULA or URA). However, there are pathological cases where $\A(\omegabf)$ can drop rank. This could occur for example, if any RIS angle differences are repeated, {\em i.e.}, $\omegabf_{RG,\ell}-\omegabf_{RH,p} = \omegabf_{RG,\ell'}-\omegabf_{RH,p'}$ for $\ell\ne\ell'$ or $p\ne p'$, or if the BS and UE arrays have identical manifolds and share an angle between the vectors $\omegabf_{BH}$ and $\omegabf_{UG}$. Although such cases occur with probability zero, $\A(\omegabf)$ will become ill-conditioned if they are approximately true, and numerical problems would ensue.} for all possible $\omegabf$, one could ignore the structure of $\gammabf_{GH}$ and use the deterministic ML (DML) criterion to estimate $\omegabf$, setting $\omegabf_{RH,1}=[0 \; \; 0]^T$ to make the model identifiable:
\beq[dml]
\hat{\omegabf} = \arg\min_\omegabf \; \mathbf{\mathfrak{y}}^H \P^\perp_{\Z\A(\omegabf)} \mathbf{\mathfrak{y}} \; ,
\eeq
where $\P^\perp_{\Z\A(\omegabf)}$ is the projection orthogonal to the columns of the effective array response $\Z\A(\omegabf)$. This would require a non-convex optimization over the $3(d_G+d_H)-2$ spatial frequencies in $\omegabf$. 

One special case worth mentioning occurs when $d_H=d_G=1$, or when the UE-RIS and RIS-BS channels are LoS. In this case, \eqref{aoagen} simplifies to $\hbf_c=\gamma \abf(\omegabf)$, where 
\beq[dhdg1]
\abf(\omegabf) = \abf_R(\omegabf_{RG}) \otimes \abf^*_U(\omega_{UG}) \otimes \abf_B(\omega_{BH}) \; ,
\eeq
with 4 angle parameters of interest: $\omega_{BH}, \omega_{UG}, \omegabf_{RG}$. This results in a standard single-snapshot AoA estimation problem, and the vector $\omegabf$ can be determined either by maximizing the beamforming criterion $|\mathbf{\mathfrak{y}}^H \Zbf \abf(\omegabf)|^2$ or minimizing the DML criterion $\mathbf{\mathfrak{y}}^H \P^\perp_{\Z\a(\omegabf)} \mathbf{\mathfrak{y}}$, and in either case setting $\hat{\gamma}=\abf^\dagger(\hat{\omegabf}) \mathbf{\mathfrak{y}}/\sqrt{P}$. Another way that beamforming can be applied in the general case with arbitrary $d_H$ and $d_G$ is to ignore the interdependence of the columns of $\A(\omegabf)$ on different combinations of the elements of $\omegabf$, and just treat the angle parameters of each column as if they were independent variables. This reparameterizes the model in~\eqref{aoagen} and~\eqref{Aok2} as
\begin{subequations}
\label{aoagen2}
\begin{align}
\mathbf{\mathfrak{y}} &= \sqrt{P} \Zbf \A(\omegabf') \gammabf + \mathbf{\mathfrak{n}} \\
\A(\omegabf') &= \left[ \abf(\omegabf'_1) \; \; \cdots \; \; \abf(\omegabf'_{d_Hd_G}) \right] \\
\abf(\omegabf'_k) &= \abf_R(\omegabf'_{k,1}) \otimes \abf^*_U(\omegabf'_{k,2}) \otimes \abf_B(\omegabf'_{k,3}) \label{aoagen2c} \; ,
\end{align}
\end{subequations}
where $\gammabf$ is an arbitrary $d_Hd_G$ vector, and $\omegabf'$ has 4 elements, one each for $\omegabf'_{k,2}$ and $\omegabf'_{k,3}$ since they are 1D spatial frequencies, and two for $\omegabf'_{k,1}$ since it is 2D. While this increases the number of angle parameters that must be estimated to $4d_Hd_G$, the beamforming criterion $\|\mathbf{\mathfrak{y}}^H \Zbf \abf(\omegabf')\|_2$ can be applied since the columns of $\A(\omegabf')$ are identically parameterized. This would require a search for $d_Hd_G$ local maxima in a 4-dimensional space. In this case, $\hat{\gammabf}'=\A^\dagger(\hat{\omegabf}') \mathbf{\mathfrak{y}}/\sqrt{P}$. 

CSI estimation for the dictionary-based model in~\eqref{csgen} also requires an unconventional approach, due to the sparsity pattern in $\qbf_{GH}$ \cite{2019Chen,ZhouPRPS21,2021dai}. The first issue is again due to the ambiguity in specifying the spatial frequencies of the RIS AoAs and AoDs; any circular shift in the columns of $\A_{RD}$ or the rows of $\Q_{RG}$ and $\Q_{RH}$ will leave $\A_D$ unchanged in~\eqref{ADdef}. As before, this ambiguity can be rectified by forcing the first column of $\Q_{RH}$ to have its non-zero element in the position corresponding to $\omegabf_{RH}=[0 \; \, 0]^T$. The second issue is most easily understood in via Eq.~\eqref{QGHdef}. The matrix $\Q_1=\left(\Q_{UG}\Gammabf^*_G \Q_{RG}^T\right)$ has $d_G$ non-zero elements at the row/column coordinates corresponding to the AoA/AoD pairs of $\G$, while $\Q_2=\left(\Q_{BH}\Gammabf_H\Q_{RH}^T\right)$ has $d_H$ non-zero elements at positions corresponding to the AoA/AoD pairs of $\H$. The Kronecker product of $\Q_1$ with $\Q_2$ then repeats the sparse structure of $\Q_2$ at the block rows/columns in $\Q_{GH}$ corresponding to the non-zero elements of $\Q_2$. In addition, the non-zero values in these repeating blocks are scaled versions of one another; the $d_H$ non-zero elements in $\Q_2$ correspond to $\gammabf_H$, and each time they are repeated in $\Q_{GH}$ they are multiplied by a different element of $\gammabf_G$. This structure in $\Q_{GH}$ in turn creates a corresponding repeated pattern in $\qbf_{GH}$. Thus, $\qbf_{GH}$ is $d_Hd_G$-sparse, but it has only $d_H+d_G-1$ unique complex entries. While cumbersome, imposing the required structured sparsity constraint on $\qbf_{GH}$ is possible \cite{2019Chen,2021dai}. 

A bigger issue with the CS approach in the general case is the size of the dictionary in~\eqref{csgen}, which has $N_{BD}N_{UD}N_{RD}^2$ elements. If we assume 100 grid points for each of the spatial frequency dimensions, this results in $10^{12}$ elements! A smaller dimensional problem can be formulated based on the parameterization in~\eqref{aoagen2}, such that
\beq[ADsmall]
\mathbf{\mathfrak{y}} = \sqrt{P} \Zbf \A'_D \qbf' + \mathbf{\mathfrak{n}} \; ,
\eeq
where $\qbf'$ is $d_Hd_G$-sparse and unstructured, and the dictionary $\Z\A'_D$ is composed of $N_{RD}N_{UD}N_{BD}$ entries defined by
\beq[ADprime]
\Z\A'_D = \Z \left( \A_{RD} \otimes \A^*_{UD} \otimes \A_{BD} \right) \; .
\eeq
While the dictionary is now a factor of $N_{RD}$ smaller, the resulting problem is likely still intractable. Assuming LoS propagation only serves to reduce the sparsity level, without reducing the dimension of the dictionary. Consequently, a more tractable approach is needed, as described next.

\subsubsection{A Simpler Decoupled Approach}\label{decouple}
In this approach, the channel estimation is decoupled into two stages; in the first, the BS and UE components of the channel are determined from an initial set of pilot data and then removed from the composite channel, and in the second additional pilot data are used to estimate the remainder of the channel \cite{2021Jiguang,2021Ardah}. In particular, stage~1 assumes that the UE transmits a $K \times T_1$ matrix of orthogonal ($T_1 \ge K$) pilot data $\Xbf_1$ while the RIS holds a fixed reflection pattern $\phibf$, which results in the following received signal at the BS:
\begin{subequations}
\begin{align}
    \Ybf_1 &= \sqrt{P} \Hbf \bar{\Phibf} \Gbf^H \Xbf_1 + \Nbf_1 \\
    &= \sqrt{P}\A_{B}(\omegabf_{BH}) \Xibf \A_U^H(\omegabf_{UG})\Xbf_1 + \Nbf_1 \label{data1b} \\
    &= \sqrt{P}\A_{BD} \Q_{BH} \Xibf \Q_{UG}^T \A^H_{UD}\Xbf_1 + \Nbf_1 \label{data1c} \; ,
\end{align}
\end{subequations}
where  
\beq[Xidef]
\Xibf = \Gammabf_H \A_R^H(\omegabf_{RH})\Phibf\A_R(\omegabf_{RG})\Gammabf^*_G \; .
\eeq
We see immediately that, assuming $M > d_H$, \eqref{data1b}-\eqref{data1c} are in the form of a standard AoA estimation problem as in~\eqref{ASN}, and thus $\omegabf_{BH}$ could be estimated from $\Y_1$ using any AoA estimation or compressive sensing algorithm\footnote{Note that in some work it is assumed that the BS-RIS channel changes slowly since the BS and RIS are in fixed locations. In such cases $\omegabf_{RH}$ can be estimated infrequently and thus may already be known, and hence the above estimation step may not always be necessary \cite{2020Ziwei,2020dai,2019hang,YouZZ20}.}. Assuming $K \ge d_G$, estimation of $\omegabf_{UG}$ can also be performed separately based on the following equation:
\begin{subequations}
\label{Aueq}
\begin{align}
\frac{\X_1 \Y_1^H}{T_1\sqrt{P}} &= \A_U(\omegabf_{UG}) \Xibf^H \A^H_B(\omegabf_{BH}) + \frac{\X_1\Nbf_1^H}{T_1\sqrt{P}} \\
&= \A_U(\omegabf_{UG}) \Sbf + \frac{1}{T_1\sqrt{P}}\X_1\Nbf_1^H \; .
\end{align}
\end{subequations}

For the second stage, the UE transmits $T_2$ pilots $\xbf_t, t=1,\cdots,T_2$, and the RIS reflection pattern also changes at the symbol rate: $\Psibf^H = \left[ \phibf_1 \; \; \cdots \; \; \phibf_{T_2} \right]$. The training received at the BS data at time $t$ can be represented as follows:
\begin{subequations}
\label{UEtd}
\begin{align}
\Y_2 &= \left[ \ybf_1 \; \; \cdots \; \; \ybf_{T_2} \right] \\
\ybf_t &= \B_t \Gammabf_{GH} \C(\omegabf_{RG},\omegabf_{RH}) \phibf_t +\nbf_t \; ,
\end{align}
\end{subequations}
where 
\beq[Bdef]
\B_t = \sqrt{P} \left( \xbf_t^T \otimes \I_M \right) \left(\A^*_U(\omegabf_{UG}) \otimes \A_B(\omegabf_{BH}) \right) \; ,
\eeq
and $\C(\omegabf_{RG},\omegabf_{RH})$ is defined in~\eqref{Cdef}. With sufficiently accurate estimates of $\omegabf_{BH}$ and $\omegabf_{UG}$ from stage~1, one can eliminate $\B_t$ from~\eqref{UEtd} and form the matrix
\begin{subequations}
\label{YBdef}
\begin{align}
\Y_B &= \left[ \B_1^\dagger \ybf_1 \; \; \cdots \; \; \B_{T_2}^\dagger \ybf_{T_2} \right]^T \\
&\simeq \Psibf^* \left(\Abf_R^T(\omegabf_{RG})  \diamond \Abf_R^H(\omegabf_{RH})\right)^T \Gammabf_{GH} + \N_B \label{YBeq2} \\
&\simeq \Psibf^* \left( \Abf_{RD}^T \diamond \Abf_{RD}^H \right)^T (\Q_{RG}\otimes\Q_{RH}) \Gammabf_{GH} +\N_B \label{YBeq3} \; ,
\end{align}
\end{subequations}
where $\N_B$ is defined similarly to $\Y_B$. 

A sparse estimation problem could be set up for $\omegabf_{RG}$ and $\omegabf_{RH}$ based on vectorizing~\eqref{YBeq3}, and while the resulting dictionary is significantly smaller than in~\eqref{csgen}, it still has $N_{RD}^2$ elements. Instead, a much simpler solution can be found by noting the structure of $\C(\omegabf_{RG},\omegabf_{RH})$ described in~\eqref{Cstruct}, which indicates that each of the $d_Gd_H$ columns of~\eqref{YBeq2} depends only on a pair of 2D angles, one from $\omegabf_{RG}$ and one from $\omegabf_{RH}$. In particular, for the $k$-th column,
\beq[Ybks]
\ybf_{B,k} \simeq \gamma_k \Psibf^* \abf_R(\omegabf_{RG,\ell}-\omegabf_{RH,p}) + \nbf_{B,k} \; ,
\eeq
where $\gamma_k$ is the $k$-th element of $\gammabf^*_G\otimes \gammabf_H$ and $k=(\ell-1)d_H+p$ for $\ell=1,\cdots,d_H$ and $p=1,\cdots,d_G$. Each of the $d_Hd_G$ columns of $\Y_B$ is thus approximately equivalent to a single snapshot from a $T_2$-element ``array'' with a single 2D spatial frequency. As before, the gains and 2D frequencies of the columns are interrelated, but if one ignores this fact, the unknown part of the channel in~\eqref{UEtd} can be reconstructed in a suboptimal way by solving a series of $d_Hd_G$ one-dimensional AoA estimation problems on each of the columns of $\Y_B$. In particular, if $\cbf_k^T$ represents the $k$-th row of $\Gammabf_{GH} \C(\omegabf_{RG},\omegabf_{RH})$, then $\cbf_k^T$ is estimated using the estimates of $\gamma_k$ and $\omegabf_k=\omegabf_{RG,\ell}-\omegabf_{RH,p}$ obtained for the $k$-th column of $\Y_B$:
\beq[ceq]
\hat{\cbf}_k = \hat{\gamma}_k \abf_R(\hat{\omegabf}_k) \; .
\eeq
For LoS channels with $d_H=d_G=1$, only a single sparse estimation problem needs to be solved.

\subsection{Wideband Single User MIMO}

As in Section~\ref{sec:unofdm}, for geometric channel models with OFDM signals, either time- or frequency domain models can be employed. We will focus on the time-domain approach here as it is a bit more straightforward. Our starting point is Eq.~\eqref{totofdm}, which differs from~\eqref{aoacsgen} in that the wideband channel parameter vector $\mathbf{\mathfrak{h}}_c$ is composed of $L$ single-tap terms like $\hbf_c$ stacked together. Partition the columns of $\mathbf{\mathfrak{Z}}$ into $L$ blocks, each block corresponding to one of the taps $\hbf_c(k)$, $k=1,\cdots,L$, in $\mathbf{\mathfrak{h}}_c$:
\beq[Zfrak]
\mathbf{\mathfrak{Z}} = \left[ \mathbf{\mathfrak{Z}}(0) \; \; \mathbf{\mathfrak{Z}}(1) \; \; \cdots \; \; \mathbf{\mathfrak{Z}}(L-1) \right] \; .
\eeq
Using~\eqref{Hcabvec}, we can write
\begin{subequations}
\label{mcarr}
\begin{align}
    \mathbf{\mathfrak{y}} &= \sqrt{P} \, \mathbf{\mathfrak{A}}(\Omegabf) 
    \left[ \begin{array}{c} \gammabf_{GH}(0) \\ \vdots \\ \gammabf_{GH}(L-1) \end{array} \right] + \mathbf{\mathfrak{n}} \\
    &= \sqrt{P} \, \mathbf{\mathfrak{A}}_D
    \left[ \begin{array}{c} \qbf_{GH}(0) \\ \vdots \\ \qbf_{GH}(L-1) \end{array} \right] + \mathbf{\mathfrak{n}} \; ,
\end{align}
\end{subequations}
where
\begin{subequations}
\begin{align}
\mathbf{\mathfrak{A}}(\Omegabf) &= \Bigl[ \mathbf{\mathfrak{Z}}(0) \A(\omegabf(0)) \; \; \cdots \; \; \mathbf{\mathfrak{Z}}(L-1) \A(\omegabf(L-1)) \Bigr] \\
\mathbf{\mathfrak{A}}_D &= \Bigl[ \mathbf{\mathfrak{Z}}(0)\A_D \; \; \cdots \; \; \mathbf{\mathfrak{Z}}(L-1)\A_D \Bigr] \; ,
\end{align}
\end{subequations}
$\Omegabf=\left[ \omegabf(0)^T \; \; \cdots \; \; \omegabf^T(L-1) \right]^T$, and where $\omegabf(k), \gammabf_{GH}(k)$ and $\qbf_{GH}(k)$ are the spatial frequencies, path gains and sparse vectors associated with the channel for tap $k$. Note that we have assumed the general case where the AoAs/AoDs are potentially different for each tap. Consequently,  
while we see from~\eqref{mcarr} that the full-scale parameterization of the problem is similar to that in~\eqref{aoacsgen}, the dictionary size and parameter dimensions are all a factor of $L$ larger. 

The decoupled approach described in Section~\ref{decouple} can be exploited to significantly reduce the required complexity. As before, we ignore the direct channel and assume an initial training period of $T_1=T_{o1}N_c$ samples from $T_{o1}$ OFDM symbols where the RIS reflection state is fixed at $\phibf$. Using~\eqref{tdmimo}, after removal of the cyclic prefix we can collect all $N_c$ samples from OFDM symbol $t$ in the matrix $\Y_{1,t}$ as follows:
\begin{subequations}
\begin{align}
\Y_{1,t} &= \sqrt{P} \Bigl[ \H(0)\Phibf\G^H(0) \; \; \cdots \; \; \H(L-1) \Phibf\G^H(L-1) \Bigr] \nonumber \\
& \qquad \qquad \qquad \qquad \times \X_{1,t}+\N_{1,t} \\
& = \sqrt{P} \A_B(\Omegabf_{BH}) \Pibf \X_{1,t}+\N_{1,t}
\end{align}
\end{subequations}
where $\X_{1,t}$ is block circulant with first block row $[\xbf_{t,1} \, \cdots \, \xbf_{t,N_c}]$, $\Omegabf_{BH}=\left[ \omegabf_{BH}(0)^T \; \cdots \; \omegabf_{BH}^T(L-1) \right]^T$, and
\begin{subequations}
\begin{align}
\A_B(\Omegabf_{BH}) &= \Bigl[ \A_B(\omegabf_{BH}(0) \; \; \cdots \; \; \A_B(\omegabf_{BH}(L-1) \Bigr] \\
\Pibf &= \text{blkdiag}\Bigl( \left\{ \Xibf(0)\A_U^H(\omegabf_{UG}(0)) \right\}_{k=0}^{L-1} \Bigr) \; ,
\end{align}
\end{subequations}
where $\Xibf(k)$ is the matrix corresponding to~\eqref{Xidef} for tap $k$. Concatenating data from the $T_{o1}$ OFDM symbols yields
\begin{subequations}
\begin{align}
\Y_1 &= \sqrt{P} \left[ \Y_{1,1} \; \; \cdots \; \; \Y_{1,T_{o1}} \right] \\
& = \sqrt{P} \underbrace{\A_B(\Omegabf_{BH})}_{M \times Ld_H} \underbrace{\Pibf \X_1}_{Ld_H \times T_1}  +\N_1 \; ,
\end{align}
\end{subequations}
where $\X_1$ and $\N_1$ are defined like $\Y_1$. Assuming $M > Ld_H$, the BS AoAs for each tap of the impulse response can be estimated using standard approaches. The AoDs at the UE can be found by noting that
\begin{align}
\frac{(\Y_1\X_1^\dagger)^H}{\sqrt{P}T_1} &= \left[ \begin{array}{c} \A_U(\omegabf_{UG}(0)) \Xibf(0) \A_B^H(\omegabf_{BH}(0)) \\ \vdots \\ \A_U(\omegabf_{UG}(L-1)) \Xibf(0) \A_B^H(\omegabf_{BH}(L-1)) \end{array} \right] \nonumber \\[5pt]
& \qquad \qquad \qquad \qquad + \frac{(\N_1\X_1^\dagger)^H}{\sqrt{P}T_1} \label{Y1X1} \; .
\end{align}
Assuming $K > d_G$, the UE angles are found by solving AoA estimation problems on the $L$ blocks in~\eqref{Y1X1}.

In Stage~2, $T_{o2}$ additional OFDM training symbols are transmitted, for a total of $T_2=T_{o2}N_c$ samples. As in~\eqref{UEtd}-\eqref{Bdef}, 
\begin{subequations}
\begin{align}
\ybf_{t,s} &= \sqrt{P}\sum_{k=0}^{L-1} \B_{t,s-k} \Gammabf_{GH}(k) \C(k) \phibf_{t,s-k} + \nbf_{t,s}\\
&= \left[ \B_{t,s} \; \; \cdots \; \; \B_{t,s-L+1} \right] \nonumber \\
& \qquad \times \left[ \begin{array}{c} \Gammabf_{GH}(0)\C(0)\phibf_{t,s} \\ \vdots \\ \Gammabf_{GH}(L-1)\C(L-1)\phibf_{t,s-L+1} \end{array} \right] + \nbf_{t,s} \nonumber \\[6pt]
\B_{t,s-k} &= \sqrt{P} \left( \xbf_{t,s-k}^T \otimes \I_M \right) \nonumber \\
& \qquad \qquad \times \Bigl(\A^*_U(\omegabf_{UG}(k)) \otimes \A_B(\omegabf_{BH}(k)) \Bigr) \; ,
\end{align}
\end{subequations}
where $\C(k)=\C(\omegabf_{RG}(k),\omegabf_{RH}(k))$. Replacing $\omegabf_{UG}(k)$ and $\omegabf_{BH}(k)$ with their estimates from Stage~1, and assuming $M \ge Ld_Hd_G$, we multiply each $\ybf_{t,s}$ on the left by the estimate of the pseudo-inverse of $\left[ \B_{t,s} \; \; \cdots \; \; \B_{t,s-L+1} \right]$, stack them together and transpose as in~\eqref{YBdef}:
\begin{subequations}
\label{YBt}
\begin{align}
\underbrace{\Y_{B,t}}_{N_c \times Ld_Hd_G} &= [ \underbrace{\Y_{B,t,0}}_{N_c \times d_Hd_G} \; \; \cdots \; \; \Y_{B,t,L-1} ] \\[5pt]
\Y_{B,t,k} &\simeq \underbrace{\Psibf_{t,k}^*}_{N_c \times N} \underbrace{\C^T(k) \Gammabf_{GH}(k)}_{N \times d_Hd_G} + \N_{B,t,k} \\[5pt]
\Psibf^*_t &= \left[ \Psibf^*_{t,0} \; \; \cdots \; \; \Psibf^*_{t,L-1} \right] \; ,
\end{align}
\end{subequations}
where $\Psibf^*_t$ is block circulant with first set of rows defined by $\Psibf^T_{t,0}=\left[ \phibf_{t,1} \; \; \cdots \; \; \phibf_{t,N_c}\right]$. Stacking the result from all $T_{o2}$ training symbols $\Y^T_B = \left[ \Y^T_{B,1} \; \; \cdots \; \; \Y^T_{B,T_{o2}} \right]$  and partitioning them into $L$ blocks of $d_Hd_G$ columns each, we have
\beq[YBmc]
\bar{\Y}_{B,k} \simeq \left[ \begin{array}{c} \Psibf_{1,k}^* \\ \vdots \\ \Psibf_{N_c,k}^* \end{array} \right] \C^T(k) \Gammabf_{GH}(k) + \N_{B,k} \; ,
\eeq
where $\bar{\Y}_{B,k}$ holds columns $k+1$ through $k+L$ of $\Y_B$. This equation is equivalent in form to~\eqref{YBeq2}, and thus the methods discussed previously can be used to solve for the remaining channel parameters for path $k$. The process is then repeated for all $L$ paths, $k=0,\cdots,L-1$.

\subsection{Single Antenna Scenarios}

\subsubsection{Single Antenna UE}\label{sec:saue}
When the UE has only a single antenna, $\omegabf_{UG}=\varnothing$ is the empty set and $\A_U(\omegabf_{UG})= \mathbf{1}^T_{d_G}$. The matrix $\A(\omegabf)$ in~\eqref{aoagen} still has $d_Hd_G$ columns, now given by
\beq[Aok3]
\left[ \A(\omegabf) \right]_{:k} = \abf_R(\omegabf_{RG,\ell}-\omegabf_{RH,p}) \otimes \abf_B(\omega_{BH,p}) \; ,
\eeq
where $\ell=\lfloor k/d_H \rfloor$ and $p=\text{mod}_{d_H}(k)$. We assume without loss of generality that $x_t=1, \forall \; t$, so~\eqref{Zdef} simplifies to
\beq[Zdef2]
\Zbf = \left[ \begin{array}{c} \phibf_1^T \otimes \I_M \\ \vdots \\ \phibf_T^T \otimes \I_M \end{array} \right] \; .
\eeq
Ignoring the structure of $\gammabf_{GH}$ in~\eqref{aoagen}, the DML criterion in~\eqref{dml} can be applied to estimate the $3d_h+2d_G-2$ spatial frequencies in $\omegabf$, which represents only a slight savings compared with the multi-antenna UE case. The simpler beamforming criterion can be used if one ignores the relationship between the columns of $\A(\omegabf)$ as in~\eqref{aoagen2}, treating them as independent vectors that are a function of three frequency variables, one for the BS and two for the RIS. This results in a search for $d_Hd_G$ local maxima in a 3D space. In the LoS case, the problem is solved by optimizing a function of three frequency variables in either the DML or beamforming approach.  

The CS-based model in~\eqref{csgen} for single-antenna UEs can also be approached in two ways. The first retains the full geometric structure of the channel, with
\begin{subequations}
\begin{align}
    \A_D &= \left( \Abf_{RD}^T \diamond \Abf_{RD}^H \right)^T \otimes \Abf_{BD} \\[5pt]
    \Q_{GH} &= \underbrace{\left( \gammabf^H_G \Q_{RG}^T\right)}_{1\times N_{RD}} \otimes \underbrace{\left(\Q_{BH}\Gammabf_H\Q_{RH}^T\right)}_{N_{BD}\times N_{RD}} \; .
\end{align}
\end{subequations}
In this case, the $d_Hd_G$-sparse vector $\qbf_{GH}=\text{vec}(\Q_{GH})$ has a similar structure as before, and the dictionary has $N_{BD}N_{RD}^2$ elements. The second approach ignores the sparse structure as in~\eqref{ADsmall}, except in this case the dictionary $\Z\A'_D=\Z(\A_{RD}\otimes\A_{BD})$ has only $N_{RD}N_{BD}$ terms. 

Further complexity reduction is possible using the decoupled approach described in Section~\ref{decouple}. In stage~1, $\omegabf_{BH}$ is estimated from $\Ybf_1$ as before, but estimation of $\omegabf_{UG}$ is not required. Stage~2 proceeds as before, but the solution is obtained in a different way. In particular, in this case we define
\begin{subequations}
\begin{align}
\ybf_t &= \sqrt{P} \A_B(\omegabf_{BH}) \left(\gammabf^H_G\otimes\Gammabf_{H}\right) \C(\omegabf_{RG},\omegabf_{RH}) \phibf_t +\nbf_t \nonumber \\
\Y_B &= \frac{1}{\sqrt{P}} \left[ \A_B^\dagger(\hat{\omegabf}_{BH}) \Y_2 \right]^T \\[5pt]
&\simeq \Psibf^* \C^T(\omegabf_{RG},\omegabf_{RH}) \left(\gammabf^*_G\otimes\Gammabf_{H}\right) + \N_B \; ,
\end{align}
\end{subequations}
and note that the $d_H$ columns of $\Y_B$ are now linear combinations of RIS array response vectors:
\begin{subequations}
\label{yblc}
\begin{align}
\ybf_{B,k} &\simeq \gamma_{H,k} \Psibf^* \sum_{n=1}^{d_G} \gammabf^*_{G,n} \abf_R(\omegabf_{RG,n}-\omegabf_{RH,k}) + \nbf_{B,k} \\
&\simeq \gamma_{H,K} \Psibf^* \text{diag}\Bigl( \A_R(\omegabf_{RG})\gammabf^*_G \Bigr) \abf_R(-\omegabf_{RH,k}) + \nbf_{B,k} ,
\end{align}
\end{subequations}
This special structure allows for a simpler solution than that required in the multi-antenna UE case considered in Section~\ref{decouple} \cite{ZhouPRPS21}. To see this, note that because of our identifiability conditions $\gamma_{H,1}=1$ and $\omegabf_{RH,1}=\zerobf$, the first column of $\Y_B$ is given by
\beq[yb1]
\ybf_{B,1} \simeq \Psibf^*\A_R(\omegabf_{RG})\gammabf^*_G + \nbf_{B,1} \; .
\eeq
It is clear that the parameters $\omegabf_{RG}$ and $\gammabf_G$ can be estimated from $\ybf_{B,1}$ using an AoA estimation such as beamforming or DML, or using a $d_G$-sparse CS algorithm. Once estimated, these parameters can be substituted into~\eqref{yblc}, and $\ybf_{B,k}$ can be used to estimate $\gamma_{H,k}$ and $\omegabf_{RH,k}$ for $k=2,\cdots,d_H$. Thus, instead of solving $d_Hd_G$ 1-sparse AoA estimation problems that ignore the underlying structure of the data as in Section~\ref{decouple}, for single-antenna UEs we can estimate the channel with one $d_G$-sparse estimation, followed by $d_H-1$ 1-sparse problems, that when combined provide estimates of $\omegabf_{RH}, \omegabf_{RG}, \gammabf_H, \gammabf_G$. 

\subsubsection{Single Antenna UE and BS}
Here there is no need for a first stage, as there are no angles to estimate at the BS or UE. Instead, we simply collect $T$ observations at the BS and stack them together in a $T\times 1$ vector $\ybf=[y_1 \; \; \cdots \; \; y_T]^T$, which yields
\begin{subequations}
\begin{align}
    \ybf &= \Psibf^* \C^T(\omegabf_{RG},\omegabf_{RH}) (\gammabf_G^*\otimes\gammabf_H) + \nbf \\
    &= \Psibf^* \sum_{k=1}^{d_G}\sum_{n=1}^{d_H} \gamma^*_{G,k}\gamma_{H,n} \abf_R\left(\omegabf_{RG,k}-\omegabf_{RH,n}\right) +\nbf \; .
\end{align}
\end{subequations}
We have an equivalent single-snapshot AoA estimation problem, although with the nonlinear Kronecker structure for the gains. Ignoring this as before, the spatial frequencies can be determined using a $2d_H+2d_G-2$ dimensional DML search, or a single $d_Hd_G$-sparse CS problem with or without application of the special row structure induced by the model.

\subsection{Multi-User Scenarios}

Multiple users in the geometric model can be accounted for by redefining $\Gbf$ as follows:
\beq[gstack]
\Gbf = \left[ \begin{array}{c} \Abf_U(\omegabf_{UG}^1) \Gammabf_{G^1} \Abf_R^H(\omegabf_{RG}^1) \\ \vdots \\ \Abf_U(\omegabf_{UG}^U) \Gammabf_{G^U} \Abf_R^H (\omegabf_{RG}^U) \end{array} \right] \equiv 
\left[ \begin{array}{c} \G^1 \\ \vdots \\ \G^U \end{array} \right] \; ,
\eeq
where the superscripts indicate the user index. UE $u$ is assumed to have $K_u$ antennas and the corresponding channel has $d_{G^u}$ propagation paths, so that $d_G=\sum_u d_{G^u}$. The composite channel and its vectorized form is given by
\begin{subequations}
\begin{align}
\H_c &= \left[ \begin{array}{c} \G^1 \diamond \H \\ \vdots \\ \G^U \diamond \H \end{array} \right] \\
\hbf_c &= \left[ \begin{array}{c} \hbf_c^1 \\ \vdots \\ \hbf_c^U \end{array} \right] = \Jbf \text{vec}(\H_c) \; ,
\end{align}
\end{subequations}
where the permutation matrix $\J$ is implicitly defined to group the blocks of $\hbf_c$ by user rather than by RIS elements. Compact expressions for the multi-user case can be found for the general data model that parallel those in~\eqref{aoacsgen}:
\begin{subequations}
\label{aoacsmu}
\begin{align}
\mathbf{\mathfrak{y}} &= \sqrt{P} \Zbf \Jbf \left[ \begin{array}{c} \A(\omegabf^1) \gammabf_{G^1H} \\ \vdots \\ \A(\omegabf^U) \gammabf_{G^UH} \end{array} \right] + \mathbf{\mathfrak{n}} \\
&=  \sqrt{P} \underbrace{\Zbf \left[ \J_1 \A(\omegabf^1) \; \; \cdots \; \; \J_U \A(\omegabf^U) \right]}_{\A(\omegabf)} \underbrace{\left[ \begin{array}{c} \gammabf_{G^1H} \\ \vdots \\ \gammabf_{G^UH} \end{array} \right]}_{\gammabf_{GH}} + \mathbf{\mathfrak{n}} \\
&= \sqrt{P} \underbrace{\Zbf \left[ \J_1 \A_D \; \; \cdots \; \; \J_U \A_D \right]}_{\text{effective dictionary}} \underbrace{\left[ \begin{array}{c} \qbf_{G^1H} \\ \vdots \\ \qbf_{G^UH} \end{array} \right]}_{\qbf_{GH}} + \mathbf{\mathfrak{n}} ,
\end{align}
\end{subequations}
where $\{\omegabf^u, \gammabf_{G^uH}, \qbf_{G^uH}\}$ are the angle parameters, channel gains, and sparse vectors for UE $u$ as in~\eqref{Hcabvec}. The block column $\J_u$ of the permutation matrix $\J=[\J_1 \; \; \cdots \; \; \J_U]$ is of dimension $MKN\times MK_uN$. General methods can be developed for simultaneous estimation of all the channel parameters based on these equations and the approaches discussed in Section~\ref{sec:sumimosc}, but as before the dimension of the resulting optimization problems is likely prohibitive except for certain simple cases.

The decoupled approach of Section~\ref{decouple} can be applied to reduce the estimation complexity. Provided that $M \ge d_Hd_G$, the BS AoAs $\omegabf_{BH}$ are estimated as before, while the equation for estimating the UE AoDs is slightly different than~\eqref{Aueq}:
\beq[x1y1]
\frac{\X_1 \Y_1^H}{T_1\sqrt{P}} = \left[ \begin{array}{ccc} \A_U(\omegabf^1_{UG}) & & \\ & \ddots & \\ & & \A_U(\omegabf^U_{UG}) \end{array} \right] \Sbf + \frac{\X_1\Nbf_1^H}{T_1\sqrt{P}} 
\eeq
where $\Sbf=\Xibf^H \A^H_B(\omegabf_{BH})$ and
\beq[Xidef2]
\Xibf^H = \Gammabf_G \left[ \begin{array}{c} \A_R^H(\omegabf^1_{RG}) \\ \vdots \\ \A_R^H(\omegabf^U_{RG}) \end{array} \right] \Phibf^*\A_R(\omegabf_{RH})\Gammabf^*_H \; .
\eeq
Assuming $K_u > d_{G^u}$, UE AoD estimates can be found separately for each user by processing different block rows of~\eqref{x1y1}. Once estimates of $\omegabf_{BH}$ and $\omegabf^u_{UG}$ are determined, we collect additional training data and proceed as in~\eqref{YBdef} where 
\begin{align}
\B_t &= \sqrt{P} \left( \xbf_t^T \otimes \I_M \right) \nonumber \\ & \; \times
\left(\left[ \begin{array}{ccc} \A_U(\omegabf^1_{UG}) & & \\ & \ddots & \\ & & \A_U(\omegabf^U_{UG}) \end{array} \right] \otimes \A_B(\omegabf_{BH}) \right) . 
\end{align}
This leads to 
\beq[newYB2]
\Y_B \simeq \Psibf^* \underbrace{\left[ \begin{array}{c} \Abf_R^T(\omegabf^1_{RG})  \diamond \Abf_R^H(\omegabf_{RH}) \\ \vdots \\ \Abf_R^T(\omegabf^U_{RG})  \diamond \Abf_R^H(\omegabf_{RH}) \end{array} \right]^T}_{\C^T\left(\omegabf^1_{RG},\cdots,\omegabf^U_{RG},\omegabf_{RH}\right)} \Gammabf_{GH} + \N_B \; .
\eeq
As before, each column of~\eqref{newYB2} involves only the difference between only one spatial frequency from $\{\omegabf^1_{RG},\cdots,\omegabf^U_{RG}\}$ and one from $\omegabf_{RH}$. There are $d_Hd_G$ such combinations, and thus the remainder of the channel parameters can be found by solving $d_Hd_G$ single spatial frequency estimation problems. Note also that when the UEs have only a single antenna, the simplification discussed in Section~\ref{sec:saue} holds, where only one $d_G$-dimensional AoA estimation followed by $d_H-1$ one-dimensional AoA estimation problems are necessary.

\subsection{Reducing the Complexity and Training Overhead}\label{sec:stred}

The methods discussed in Section~\ref{sec:reduce} can be used to further reduce the training required even for geometric channel models. The availability of prior knowledge of low-rank channel covariance matrices is useful for geometric models, although less for reducing the amount of training than for designing the pilot symbols $\xbf_t$ and RIS training $\phibf_t$ to improve the received SNR. Also, the common channel $\Hbf$ associated with multiple UE antennas can be exploited as before to reduce the algorithm complexity. For geometric channel models, this approach could be implemented as follows \cite{ZhouPRPS21}:
\begin{enumerate}
    \item Choose one antenna from one of the UEs, and transmit training data while the other antennas are silent to estimate the channel $\Hbf\text{diag}(\gbf_1)$.
    \item Transmit training data from the remaining UE antennas, and estimate the UE AoDs $\omegabf^k_{UG}$ as in~\eqref{Aueq} if there is only one user with multiple antennas, or as in~\eqref{x1y1} if there are multiple multi-antennas users. 
    \item Set the RIS reflection vector to a fixed value $\bar{\phibf}$ and send at least $d_G$ additional training symbols $\xbf_t$ to obtain
    \beq[redux]
    \Ybf = \Hbf \bar{\Phibf} \A_R(\omegabf_{RG})\Gammabf_G \A_U^H(\omegabf_{UG}) \Xbf + \Nbf \; .
    \eeq
    Then, multiply on the right by $\left(\A_U^H(\hat{\omegabf}_{UG}) \Xbf\right)^\dagger$. Since we have an estimate of $\Hbf$, the resulting matrix is approximately in the standard form for AoA estimation, using either steering vectors drawn from $\hat{\Hbf}\bar{\Phibf}\A_R(\omegabf_{RG})$, or an overcomplete dictionary $\hat{\Hbf}\bar{\Phibf}\A_{RD}$.
\end{enumerate}

Use of the geometric model offers a further opportunity for dramatic reductions in both computational complexity and training overhead. The AoAs and AoDs in the geometric model change very slowly, and can be considered to be stationary over multiple coherence blocks. Only the complex gains $\gammabf_G$ and $\gammabf_H$ change substantially from block to block. This suggests that, once estimated, the spatial frequencies for subsequent blocks can be considered constant, and only the gains need to be re-estimated \cite{ZhouPRPS21}. For CS-based methods, this means that the support of the sparse solution is already known. Consequently, with at least $d_Hd_G$ training samples, the channel gains in equations~\eqref{aoacsgen}, \eqref{mcarr}, or~\eqref{aoacsmu} can be estimated using least-squares; {\em e.g.,} for~\eqref{aoagen} we have
\beq[fixa]
\hat{\gammabf}_{GH} = \frac{1}{\sqrt{P}} \Bigl( \Zbf \Abf(\hat{\omegabf}) \Bigl)^\dagger \mathbf{\mathfrak{y}} \; .
\eeq

\section{Numerical Examples}\label{sec:crb}
To illustrate the CSI estimation performance for typical RIS-based scenarios, we will use the CRB rather than plotting the results of individual algorithms (for which there are too many examples to fairly consider). In particular, we will show numerical results for the CRB of $\breve{\hbf}_c^T = \left[ \text{Re}(\hbf_c)^T \; \; \text{Im}(\hbf_c)^T \right]$, where $\text{Re}(\hbf_c)$ and $\text{Im}(\hbf_c)$ respectively represent the real and imaginary parts of $\hbf_c$. 

In the unstructured case, the parameters are the elements of the channel itself: $\etabf=\breve{\hbf}_c$. Assuming temporally and spatially white Gaussian noise with variance $\sigma^2$ as in~\eqref{sumimo} and~\eqref{fulltrain}, the log-likelihood is given by 
\beq[loglike]
f_u(\breve{\hbf}_c) = -MT \ln (\pi \sigma^2) - \frac{1}{\sigma^2} \left\| \mathbf{\mathfrak{y}} - \sqrt{P} \Zbf \hbf_c \right\|^2 \; ,
\eeq
where the subscript $u$ denotes ``unstructured.'' The CRB is defined in terms of the Fisher Information Matrix (FIM):
\begin{align}
    \text{CRB}_u(\breve{\hbf}_c) &= \text{FIM}_u^{-1}(\breve{\hbf}_c) \\
    \text{FIM}_u(\breve{\hbf}_c) &= \mathbb{E} \left\{ \frac{\partial f_u(\breve{\hbf}_c)}{\partial \breve{\hbf}_c} \left( \frac{\partial f_u(\breve{\hbf}_c)}{\partial \breve{\hbf}_c} \right)^T \right\} \; ,
\end{align}
where $\mathbb{E}(\cdot)$ denotes expectation with respect to the noise distribution. It is straightforward to show that the CRB for the unstructured model is given by
\beq[crbus]
\text{CRB}_u(\breve{\hbf}_c) = \frac{\sigma^2}{2 P} \left( \breve{\Zbf}^T \breve{\Zbf} \right)^{-1} \; ,
\eeq
where
\beq[zbrev]
\breve{\Zbf} = \left[ \begin{array}{cr} \text{Re}(\Zbf) & -\text{Im}(\Zbf) \\ \text{Im}(\Zbf) & \text{Re}(\Zbf) \end{array} \right]
\eeq
and $\Zbf$ is as defined in~\eqref{fulltrain}. If $\Zbf$ is designed to be orthogonal as in Section~\ref{sec:ls}, then the CRB simplifies to
\beq[crbus2]
\text{CRB}_u(\breve{\hbf}_c) = \frac{\sigma^2}{2 P T} \I_{2MK(N+1)} \; ,
\eeq
signifying that the lower bound is identical for every element of $\breve{\hbf}_c$ (note that we have assumed the RIS element gains are $\beta=1$). Note that the CRB in the unstructured case is independent of the number of BS antennas, RIS elements, and UEs (although $T\ge K(N+1)$ must hold for the model to be identifiable, and hence for the FIM to be invertible).

The CRB for the geometric channel model depends on the parameter vector $\etabf$ defined by the angles and gains listed in Table~\ref{tab:param}. The CRB for $\etabf$ is more difficult to compute, requiring a large number of tedious derivative calculations which we do not include here. When one forms an estimate of the composite channel $\hbf_c$ from the estimate of the parameters in $\etabf$, then the resulting CRB for the channel is given by
\beq[crbgeo]
\text{CRB}_s(\breve{\hbf}_c) = \frac{\partial \breve{\hbf}_c}{\partial \etabf} \text{CRB}(\etabf) \left( \frac{\partial \breve{\hbf}_c}{\partial \etabf} \right)^T \; ,
\eeq
where here the subscript $s$ is for ``structured.'' The examples below compare $\text{CRB}_u(\breve{\hbf}_c)$ and $\text{CRB}_s(\breve{\hbf}_c)$ for several different scenarios. The performance metric adopted in all numerical examples is the average of the diagonal elements of the CRB over a large number of different (geometric) channel realizations with random AoAs/AoDs and path gains. 

The first case considers a single-UE with $K \in \{1,2\}$ antennas, orthogonal pilot signals, and uniformly distributed RIS phases during training. The AoAs/AoDs were generated using a uniform distribution over $[-90^\circ,90^\circ]$ for azimuth and $[0^\circ,90^\circ]$ for elevation. The path gains were generated as unit-variance Rayleigh random variables. Since the path gains are independent of the number of paths, a channel with more paths will generally make a larger contribution to the overall SNR. The path gain distribution is the same for all three channels $\Hbf_d, \H, \G$, which corresponds to a case where the BS, RIS and UE are approximately located at the vertices of an equilateral triangle with similar propagation characteristics. Also, this assumption means that, even though during channel estimation the RIS phases $\Phibf_t$ are not chosen to maximize the coherent gain offered by the RIS, the RIS provides sufficient combining gain such that $\| \H\Phibf_t\G\| \gg \|\H_d\|$. The SNR in the examples below is defined as $\P/\sigma^2$, but the effective SNR will be a function of the channel gain $\|\H_d + \H \Phibf_t \G\|$, which will change as a function of the number of antennas, RIS elements, and path gains.

\begin{figure}
\centering
  \includegraphics[width=3.5in]{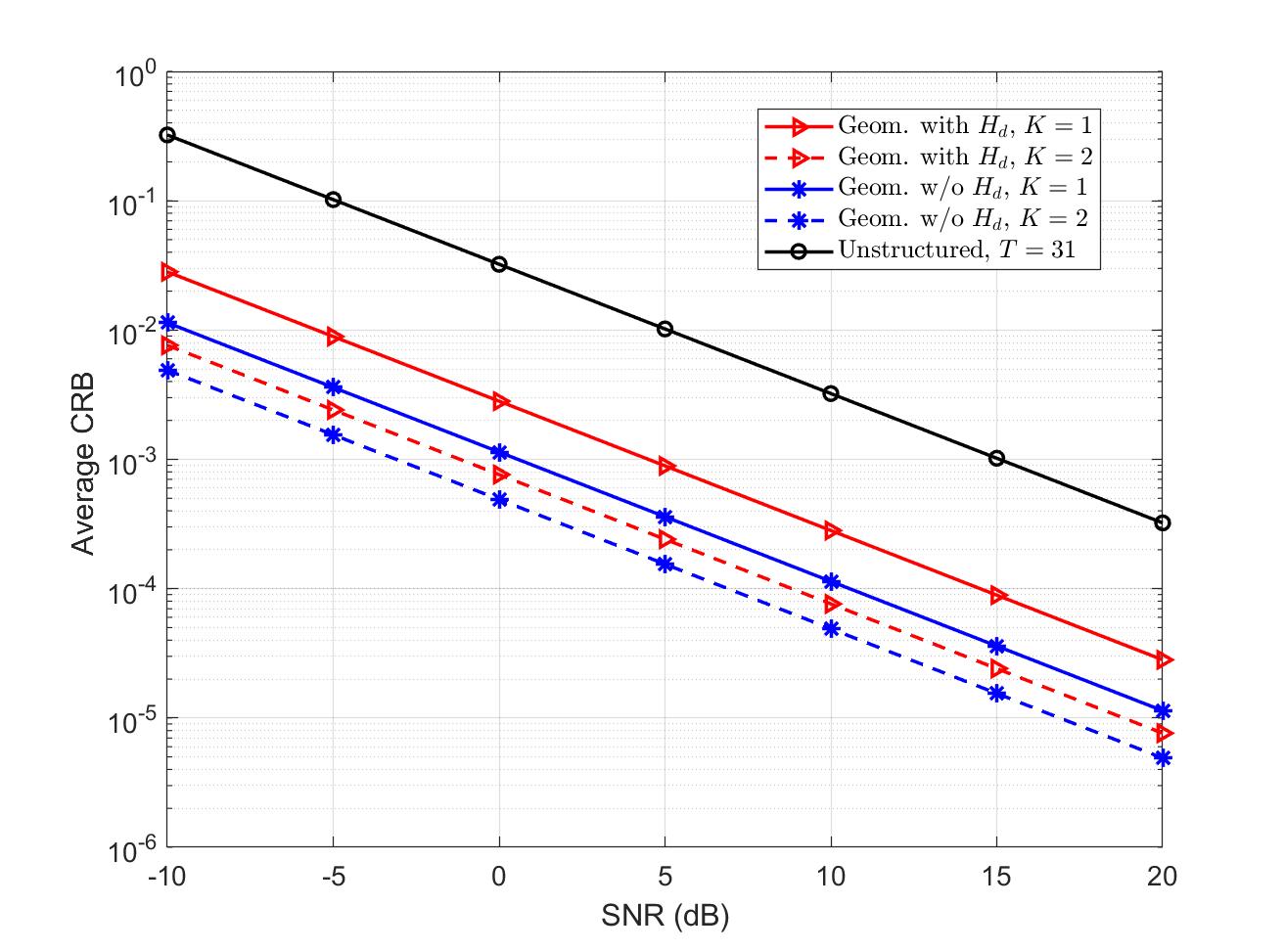} 
\caption{CRB for various channel models vs.~SNR. Channel parameters are $M=30, N=30, T=31, K=2, d_F=5, d_G=5, d_H=2$.}
\label{SGig1}
\vskip -0.5cm
\end{figure}

\begin{figure}
\centering
  \includegraphics[width=3.5in]{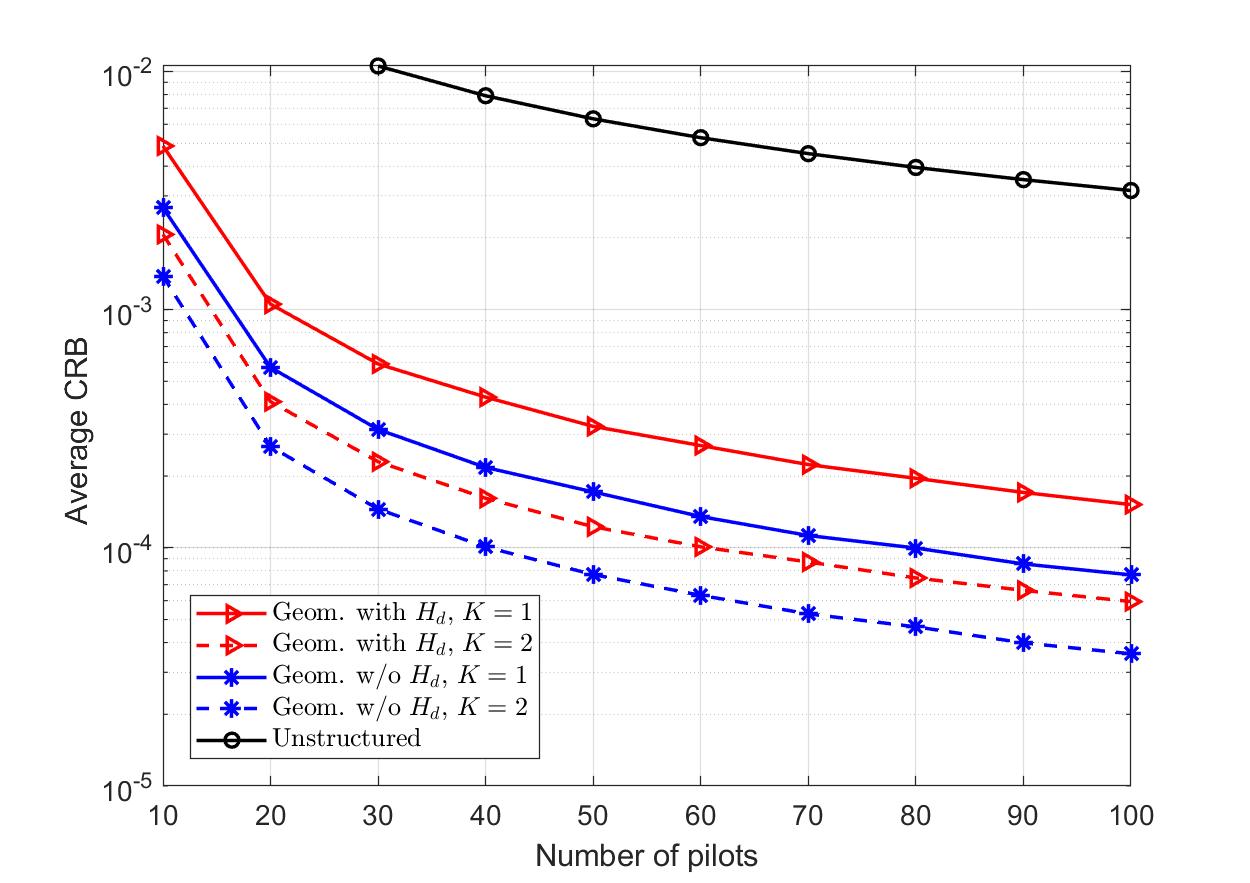} 
\caption{CRB for various channel models vs.~number of training samples $T$. Channel parameters are $M=30, N=30, \text{SNR}=5\text{dB}, K=2, d_F=5, d_G=5, d_H=2$.}
\label{SGig2}
\vskip -0.5cm
\end{figure}

\begin{figure}
\centering
  \includegraphics[width=3.5in]{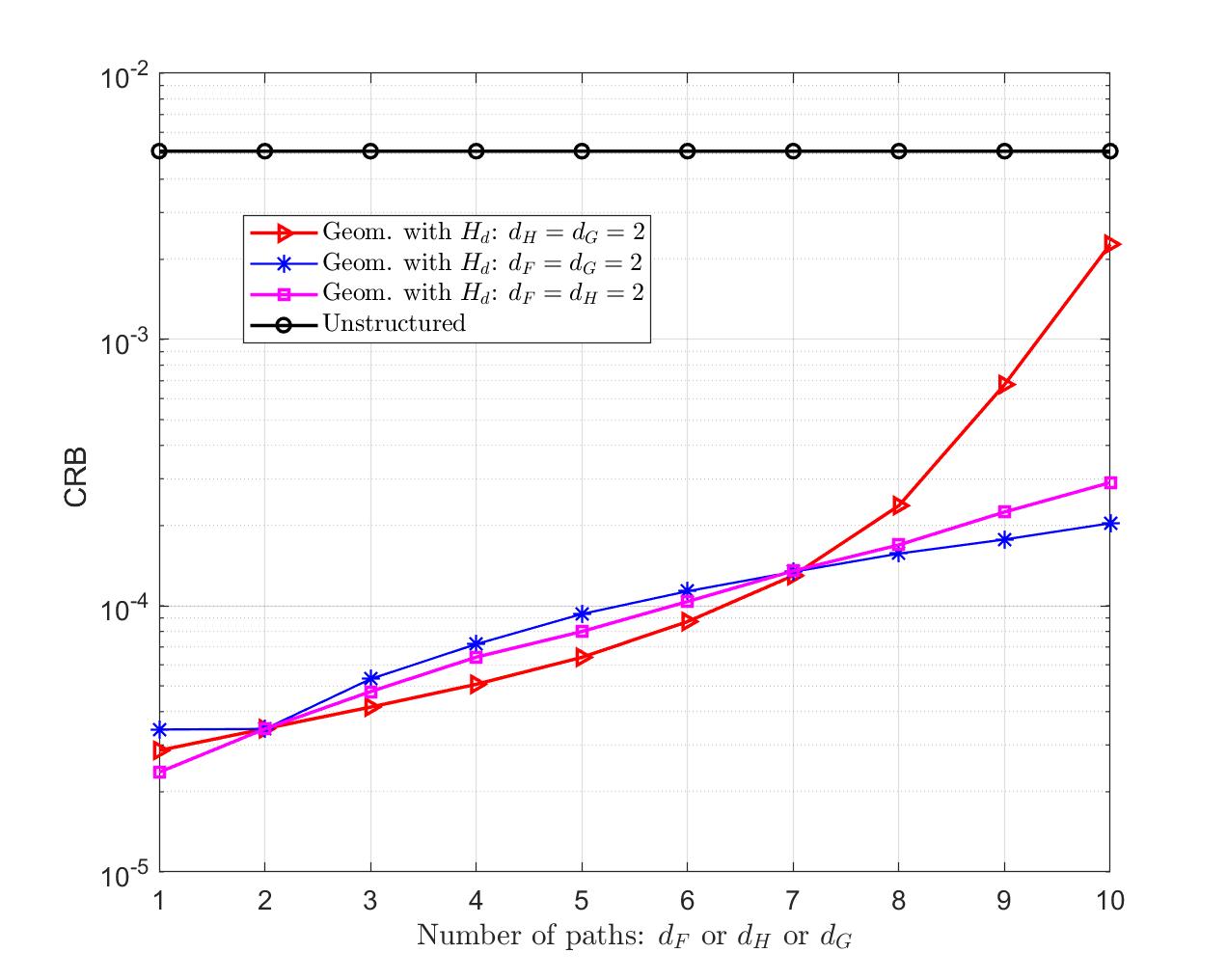} 
\caption{CRB for various models vs.~number of propagation paths. Channel parameters are $M=30, N=30, T=62, K=2, \text{SNR}=5\text{dB}$.}
\label{SGig4}
\vskip -0.5cm
\end{figure}

Figs.~\ref{SGig1}-\ref{SGig4} assume a BS with $M=30$ antennas and an RIS with $N=30$ elements. Fig.~\ref{SGig1} plots the CRB as a function of SNR for a case with $d_H=2$, $d_F=d_G=5$, and $T=31$ training samples, which is the minimum required for the case of a single-antenna UE ($K=1$). The CRB for the unstructured model and the geometric model with and without the direct BS-UE channel $\H_d$ are included. The advantage of the geometric model is clear in this example, providing 10-15dB of gain over the unstructured case, although a few dB of this gain would be lost due to imprecise BS and RIS array calibration \cite{RaoMS19}. 

The advantage of the geometric model in terms of training data is apparent from Fig.~\ref{SGig2} for the same scenario with an SNR of 5dB; we see that the same performance can be achieved with more than order of magnitude fewer training samples. Note that the plots show a degradation in geometric channel estimation performance when the direct channel is present; this is due to the observation above that for the assumed model, we have $\| \H\Phibf_t\G\| \gg \|\H_d\|$. In effect, when $\H_d$ is present, we are faced with the task of estimating additional parameters that are observed more weakly in the data.

Fig.~\ref{SGig4} plots the CRB as a function of the number of paths present in the geometric channel model, again assuming $M=N=30$ and the minimum number of $T=K(N+1)=62$ training samples. The red curve assumes $d_H=d_G=2$ and shows an increase in the CRB as $d_F$ increases, the blue curve assumes $d_F=d_G=2$ and plots versus $d_H$, while the magenta curve is for $d_F=d_H=2$ versus $d_G$. While the presence of more paths increases the power of the received signals in this example, this gain is offset by the fact that more parameters must be estimated, and hence the geometric CRB increases with the number of paths. The increase due to larger values of $d_F$ causes the most dramatic increase for the reason noted in the previous paragraph. For the common case where the direct channel is blocked, it is clear that the geometric model can include a relatively large number of paths before its performance degrades to the level of the unstructured case.

\begin{figure}
\centering
 \includegraphics[width=3.5in]{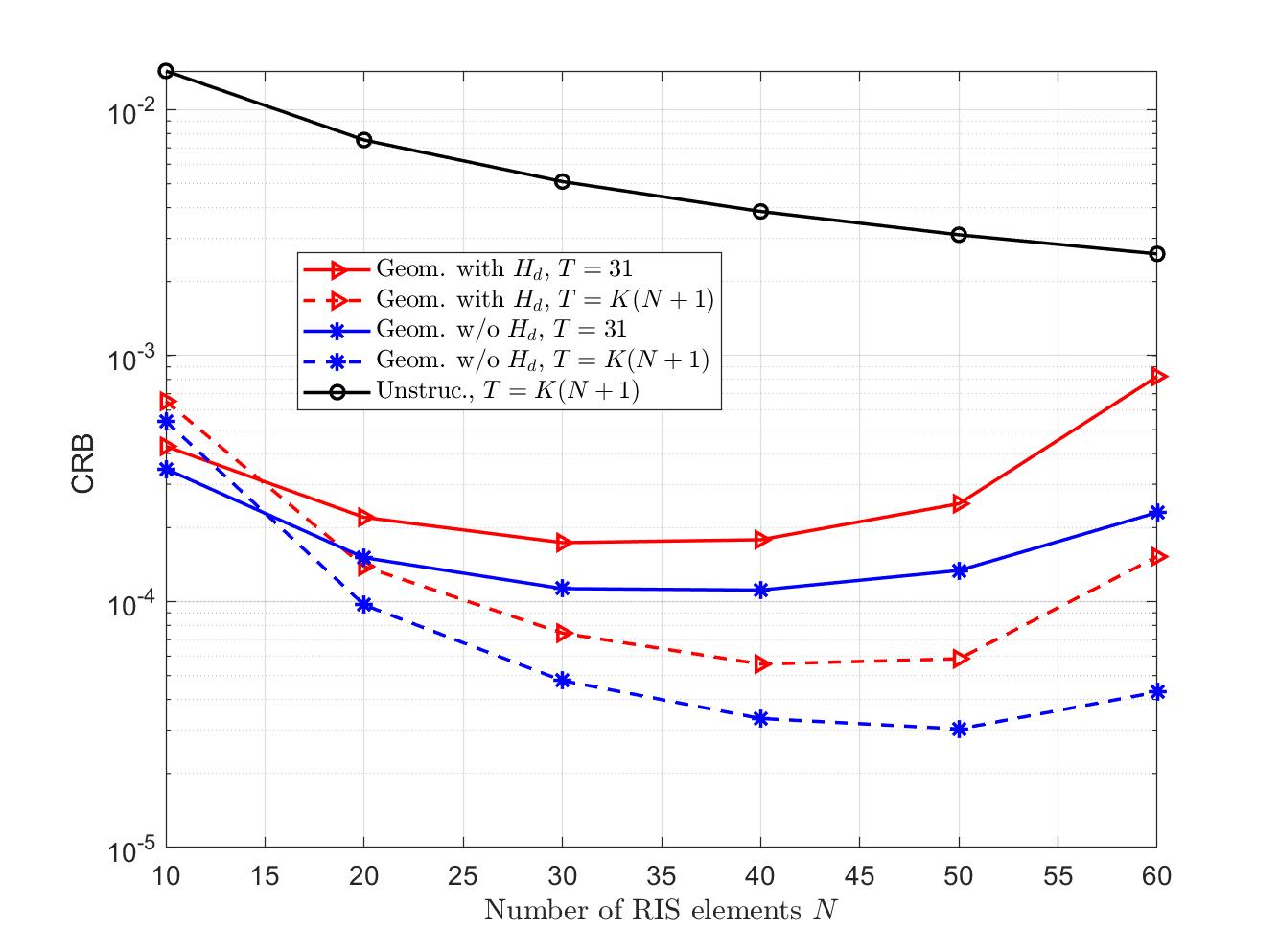} 
\caption{CRB for various channel models vs.~number of RIS elements $N$, where $M+N=70$. Channel parameters are $\text{SNR}=5\text{dB}, K=2, d_F=5, d_G=5, d_H=2$. For the unstructured model and the dashed blue and red curves (geometric model), we have $T=K(N+1)$. For the solid blue and red curves (geometric model), $T=31$.}
\label{SGig8}
\vskip -0.5cm
\end{figure}

The last example for the first scenario is depicted in Fig.~\ref{SGig8}, which shows the achievable channel estimation performance as a function of $N$ assuming that $M+N=70$, with the other parameters set as before ($K=2, d_F=5, d_G=5, d_H=2, \text{SNR}=5\text{dB}$). The curve for the unstructured model assumes that $T$ is increasing with $N$ according to $T=K(N+1)$, which explains why the performance improves as the size of the RIS increases. The dashed curves for the geometric model assume the same increasing values for $T$, while the solid lines assume a fixed value of $T=31$. For fixed $T$, the best performance is achieved when $M \simeq N$, while larger values of $T$ favor the use of a larger RIS and a smaller number of BS antennas.

\begin{figure}
\centering
  \includegraphics[width=3.5in]{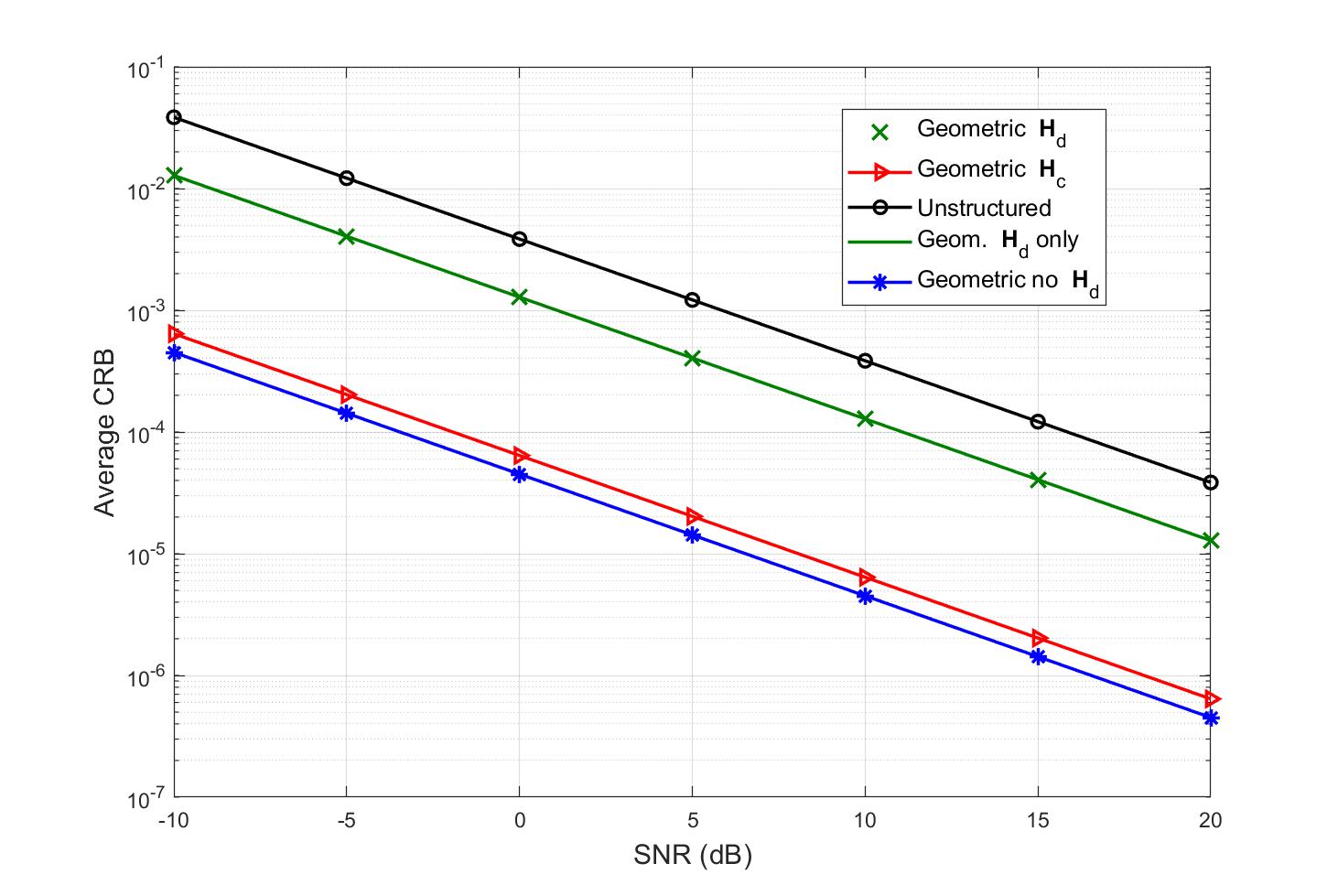} 
\caption{CRB for various channel models vs.~SNR. Channel parameters are $M=6, N=64, T=260, U=2, K=4, d_F=2, d_G=3, d_H=2$.}
\label{Sfig1}
\vskip -0.5cm
\end{figure}

\begin{figure}
\centering
  \includegraphics[width=3.5in]{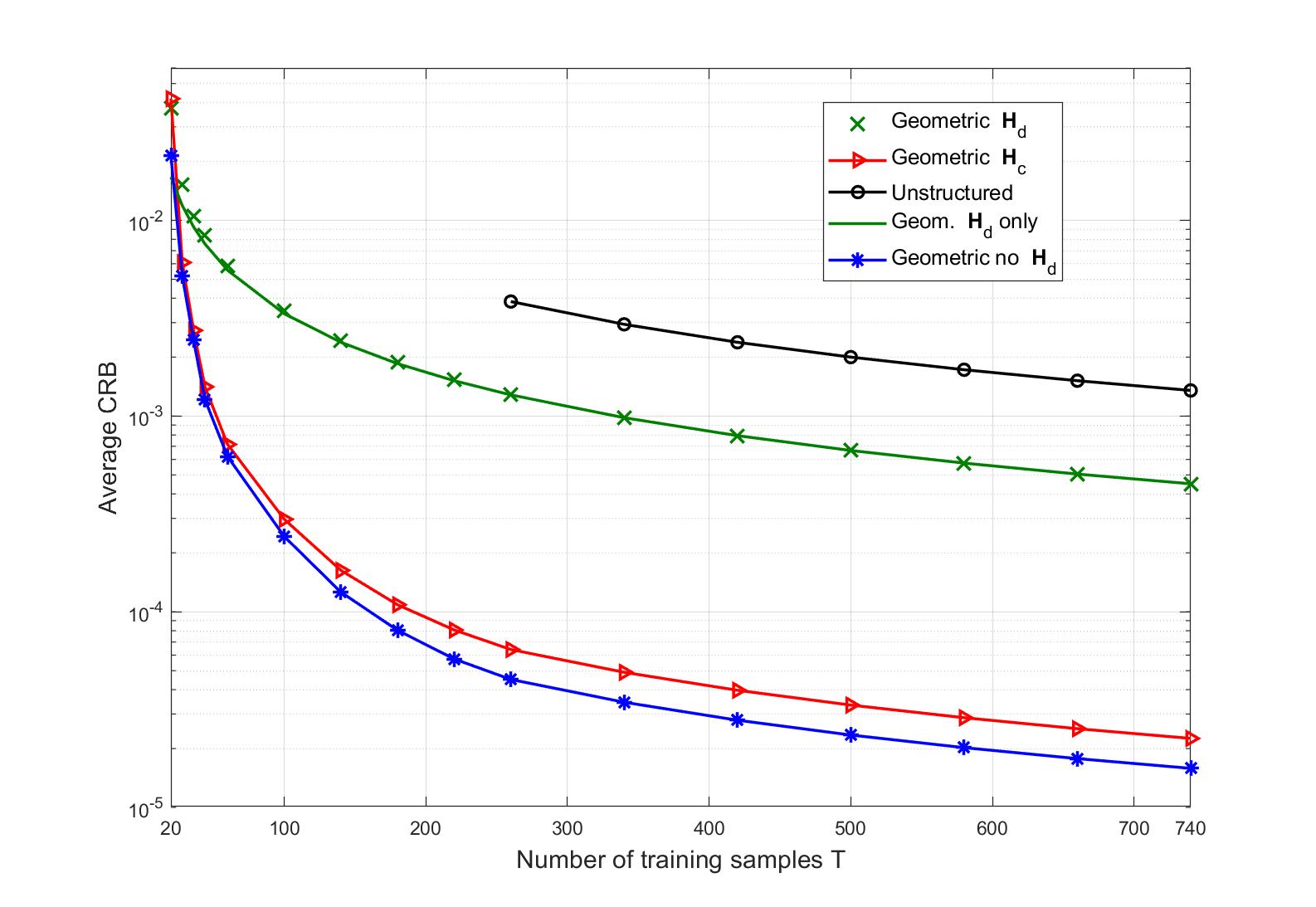} 
\caption{CRB for various channel models vs.~number of training samples $T$. Channel parameters are $M=6, N=64, \text{SNR}=0\text{dB}, U=2, K=4, d_F=2, d_G=3, d_H=2$.}
\label{Sfig2}
\vskip -0.5cm
\end{figure}

\begin{figure}
\centering
  \includegraphics[width=3.5in]{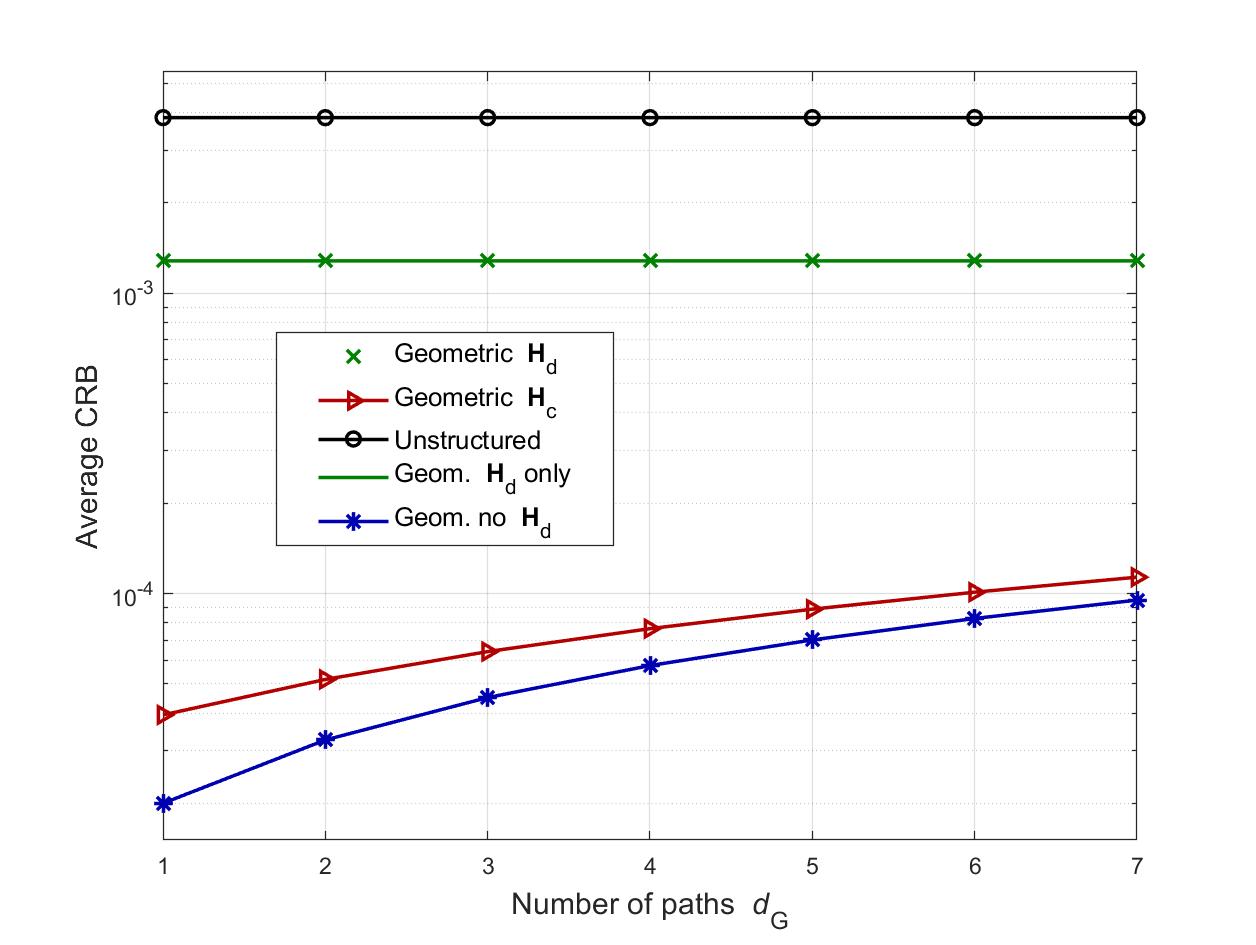} 
\caption{CRB for various models vs.~number of propagation paths $d_G$. Channel parameters are $M=6, N=64, T=260, U=2, K=4, \text{SNR}=0\text{dB}, d_F=2, d_H=2$.}
\label{Sfig5}
\vskip -0.5cm
\end{figure}

The second scenario is different from the first in a number of ways, but the conclusions are essentially the same. Unlike the previous case, there are $U=2$ users with 2 antennas each ($K=4$), the RIS training sequences are not random but rather chosen such that the diagonal CRB in~\eqref{crbus2} holds, the BS has $M=6$ antennas, and there are $d_H=d_F=2$ paths in the BS-RIS and BS-UE channels. Except for Fig.~\ref{Sfig5}, $d_G=3$. In addition, the Rayleigh distributed path gains in this example have a variance equal to the reciprocal of the number of paths. Figs.~\ref{Sfig1}-\ref{Sfig5} assume a larger RIS with $N=64$ elements, and thus the LS channel estimator requires at least $T=K(N+1)=260$ training samples in order to obtain a unique estimate. This value for $T$ will be assumed, except for the case where performance is plotted versus $T$. 

\begin{figure}
\centering
  \includegraphics[width=3.5in]{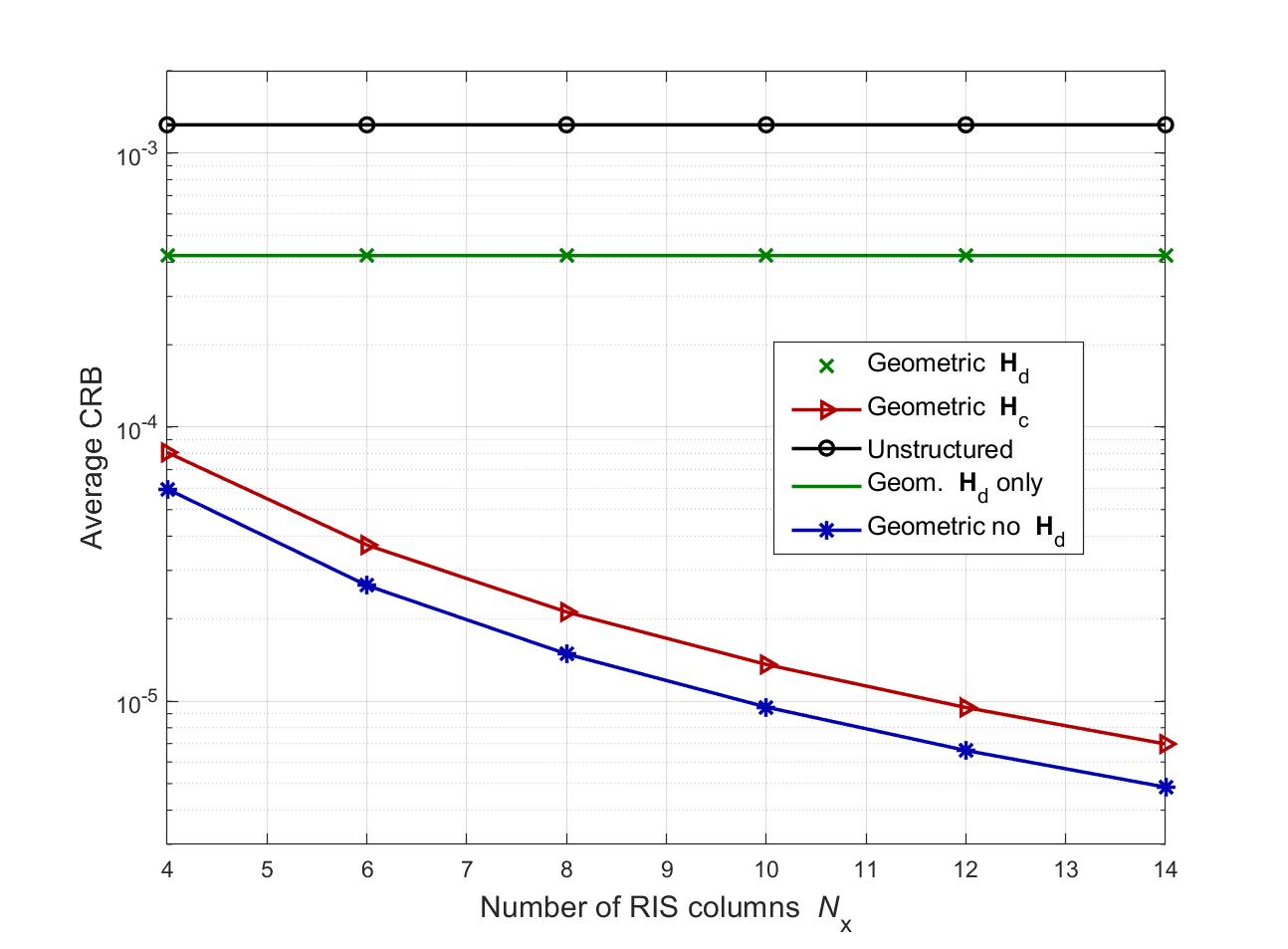} 
\caption{CRB for various channel models vs.~number of columns $N_x$ in a square RIS with $N=N_x^2$ elements. Channel parameters are $M=6, T=788, \text{SNR}=0\text{dB}, U=2, K=4, d_F=2, d_G=3, d_H=2$.}
\label{Sfig4}
\vskip -0.5cm
\end{figure}

Figs.~\ref{Sfig1}-\ref{Sfig4} each show the CRB for five different cases: (1) ``Geometric $\H_d$'' - geometric model for $\H_d$ channel with the RIS channel present, (2) ``Geometric $\H_c$'' - geometric model for $\H_c$ with $\H_d$ included, (3) ``Unstructured'' - unstructured model (independent of $\H_d$), (4) ``Geom. $\H_d$ only'' - geometric model when only $\H_d$ is present, and (5) ``Geometric no $\H_d$'' - geometric model for $\H_c$ without $\H_d$ present. In all cases, the results for ``Geometric $\H_d$'' and ``Geom. $\H_d$ only'' are essentially identical, indicating that the presence or absence of the RIS should not impact the quality of the estimate of $\H_d$.

Fig.~\ref{Sfig1} shows the CRB versus the SNR, and as in the previous case we see an estimation gain of about 15dB for the geometric model in the ideal case. Achievable performance versus $T$ is illustrated in Fig.~\ref{Sfig2} for 0dB SNR. The curve for the unstructured case is only shown for $T \ge 260$ where the channel is identifiable, and as before there is at least an order of magnitude reduction in training data required for the geometric model. Fig.~\ref{Sfig5} provides results versus $d_G$ for 0dB SNR, and we still see a considerable gap between the achievable performance of the unstructured and geometric models even as the number of propagation paths grows larger. The CRB versus the number of RIS elements $N=N_x^2$ for 0dB SNR is shown in Fig.~\ref{Sfig4} with $T=K(N+1)=788$ in order for the unstructured model to be identifiable for the case of the largest RIS. 

\section{Additional Topics}\label{sec:other}


\subsection{Active RIS Elements}
The methods discussed above have assumed a purely passive RIS with elements that are capable of applying only a controllable phase shift to the impinging signal, together with a phase- and frequency-dependent attenuation factor. Recently, several research teams have studied the added flexibility that can be provided if the RIS is equipped with a few active receivers and local baseband processing, suggesting that there are benefits that outweigh the resulting increase in power consumption at the RIS. Most of this work has assumed only active receiving and not active transmission at the RIS, hence the additional required power is still significantly less than a similarly-sized active relay. Instead of the lower SNR data at the BS that has propagated from the UEs through the RIS, the data collected by the RIS has higher SNR due to lower path loss and will lead to higher fidelity channel estimates. In addition, the channels $\H$ and $\G$ can be separately identified, which as we have seen above is useful in cases where one component varies less rapidly than the other, and need not be estimated as frequently.

The active RIS elements are used to estimate the gains and angles of the geometric channel model, and these estimates are then used to infer the full $\H$ and $\G$ matrices using the known RIS geometry \cite{HuZZ21,TahaAA21,ChenSY21} or a deep neural network (DNN) \cite{JinZZ21}. This matrix completion approach is similar to methods proposed for channel estimation in hybrid digital/analog systems. A tensor completion approach is proposed in \cite{LinJM21}, and also extended to the wideband OFDM case. The active elements can either be embedded in fixed positions with a separate RF chain for each as in the methods above, or different linear combinations of the RIS outputs can be combined together through a single RF chain over the training interval to achieve a similar result \cite{AlexV20}. We also mention here the related work of \cite{GuanWZ21} which, instead of active RIS elements, proposes deploying active anchor nodes in known locations near the UEs to reduce the CSI estimation complexity as discussed in Section~\ref{sec:stred} and resolve the ambiguity of the composite channel.

\subsection{Double RIS Systems}
We have only considered communication links between a BS and UEs that include a single RIS, but channel estimation for scenarios with two RIS have been studied for certain special cases. In \cite{2021Changsheng}, a double RIS scenario is considered where a BS communicates with a single UE over an indirect channel that passes from the BS to RIS 1 to RIS 2 to the UE, assuming all other channels are blocked. They propose an unstructured LS channel estimator that requires at least $N_1N_2$ training samples, assuming the two RIS have $N_1$ and $N_2$ elements respectively. They also consider the special case of an LoS channel between the two RIS, which reduces the training overhead to $N_1+N_2$. A more general scenario is considered in \cite{2020Beixiongtcom}, in which both RIS have unblocked channels with the BS and UEs, in addition to the direct link between the two RIS. A two-stage unstructured LS approach is proposed where, in the first, the phase shifts of RIS 1 are held fixed while those of RIS 2 are varied, and the composite single-bounce (UE-IRS 2-BS) and double-bounce (UE-IRS 2-IRS 1-BS) channels are estimated. In stage 2, the phase of both RIS are varied in order to estimate the corresponding components of RIS 1, which are superimposed on the composite channels estimated in stage 1. Assuming $K$ single-antenna UEs, the required training overhead for this approach is of the following order:
\begin{equation}
    T = O\Bigl( \frac{KN_1N_2}{M} + K(N_1 + N_2) \Bigr) \; ,
\end{equation}
which can be large when the number of BS antennas $M$ is small. An earlier approach to the double-RIS problem in \cite{HanZDZ20} assumed LoS propagation between two RIS that both possess active receivers, which allowed the individual channel components to be determined. It is unclear if a geometric model could be used for a purely passive double-RIS scenario to achieve an identifiable parameterization.

\subsection{Learning-Based Methods}
The application of machine learning to RIS systems and channel estimation has also been growing recently. For example, the approach in \cite{ElbirPKC20} designs a DNN that directly takes the received training data and produces the channel estimate. The DNN is trained using synthetic uplink data generated assuming the direct and composite channels are Rayleigh fading. As has been noted in other work, the performance improves if not only the real and imaginary parts of the input data are provided, but also a third component related to the data, in this case its magnitude. Other work has employed the phase of the data as the third component \cite{2020Ahmet}. However, \cite{2020Ahmet} uses a federated learning approach in the downlink, in which the UEs generate local channel estimates using a learning-based optimization, and the gradients of the local networks are fed back to the BS to update the global model.

A common learning-based approach for RIS systems involves denoising an initial channel estimate with a DNN. The initial estimate can be found using the methods described above, such as LS \cite{LiuLNY21,KunduM21} or compressive sensing in the geometric model \cite{LiuG20}. Multi-stage DNN denoising networks are proposed in \cite{WangLS21,GaoDP21}, where separate DNNs are used to first denoise the direct channel (if present) with the RIS switched off, then to denoise a reduced-dimension version of the composite channel with only a subset of the RIS elements active, and finally to map the sampled composite channel to its full dimensions. A similar technique is proposed in \cite{JinZZ21} that uses a DNN to map the channel estimated with a few active RIS elements to the full RIS array. The approach in \cite{TahaAA21} also assumes the availability of active RIS receivers, but instead of attempting to estimate the channel with the full RIS, the sampled channels from the active RIS subarray are used as ``environment descriptors'' during training to find the best codebook of RIS phases that yields the highest sum rate.

The primary advantage of the above learning-based methods is that, once the neural networks are trained, the channel estimates are obtained with relatively little computation. However, since the training is performed using synthetic data, the ability of the network to handle non-idealities not present in the simulated training data is still an open question.

\section{Conclusions}\label{sec:conc}
Wireless channel estimation for RIS-based systems provides a rich source of interesting research problems. We have highlighted some of the solutions to these problems for two general classes of channel models: unstructured models that make no assumptions about the propagation environment or the RIS or array geometries, and structured or geometric models that rely on the assumption of sparse propagation paths and calibrated antenna arrays and RIS element responses. Algorithms for estimating unstructured channels are conceptually simple and robust, but for RIS-based systems they require a large training overhead and their achievable accuracy is limited due to the large number of channel coefficients to be estimated. On the other hand, geometric channel models lead to estimation of many fewer parameters and hence have a much smaller training burden and can achieve dramatically better performance. However, this improvement comes at the cost of increased algorithm complexity, a requirement to determine the model order ({\em i.e.,} number of propagation paths), and some of the performance gain will be lost in practice due to inevitable modeling errors. We have highlighted these issues using both theoretical derivations and numerical CRB examples, and we have also briefly discussed on-going research related to RIS with active elements, double-RIS systems, and machine learning algorithms. In addition to these areas, many open research problems remain, including how to estimate channels for multiple RIS when they are visible to multiple basestations and their reflections interact, how to take into account more realistic models of RIS behavior (coupled dependence of gains and phases over frequency, mutual coupling, etc.), how to reduce the complexity of compressive-sensing based approaches that normally require huge dictionaries for RIS-based CSI estimation, how to exploit available side information such as knowledge of the local propagation environment or known channel statistics for optimal design of pilots and RIS phases during training, etc.

\bibliographystyle{IEEEtran}
\bibliography{IEEEabrv,bibJournalList,bibfile}

\vspace{-32 mm}

\begin{IEEEbiography}[{\includegraphics[width=1in,height=1.25in,clip,keepaspectratio]{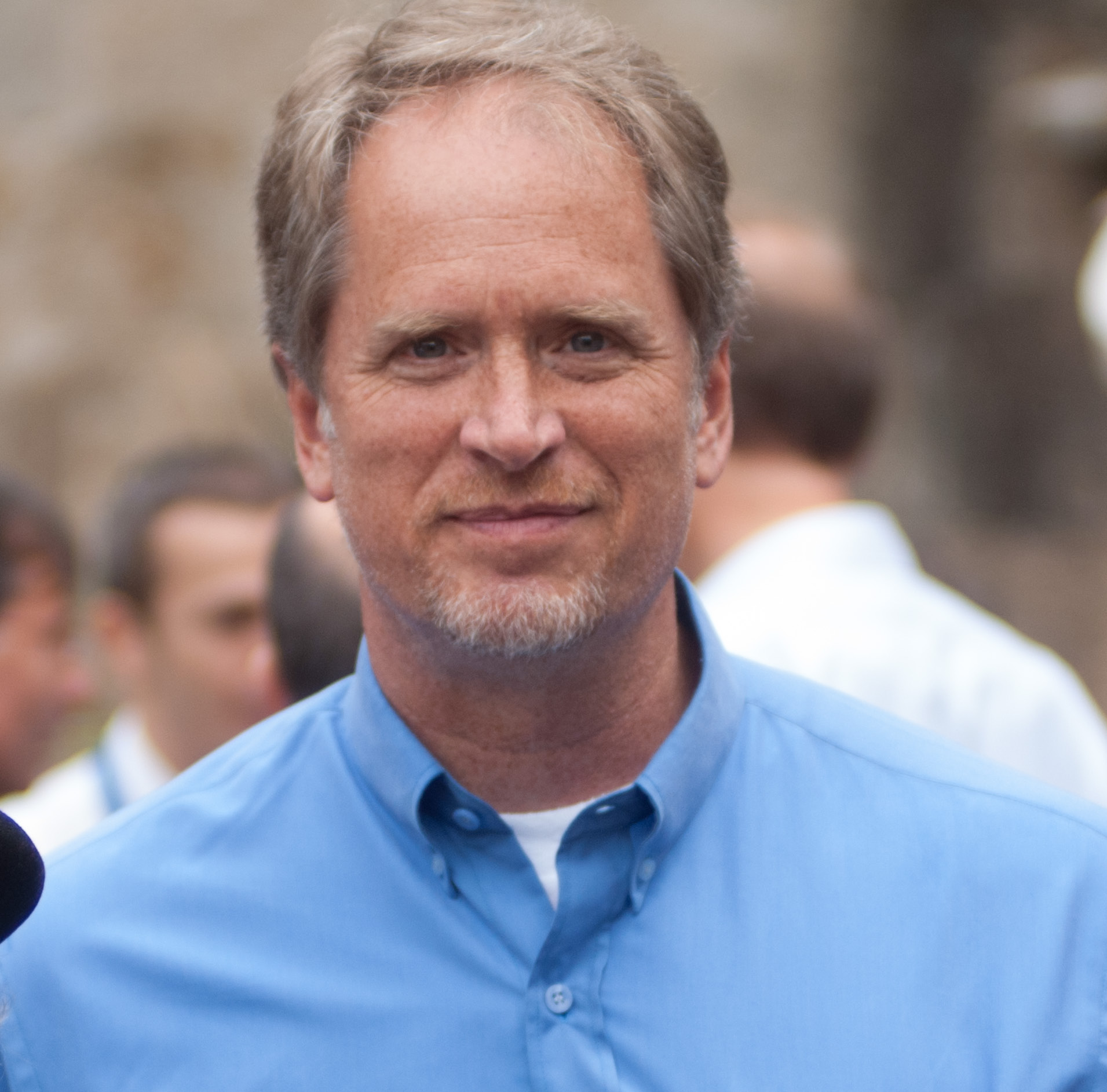}}]{A. Lee Swindlehurst} (Fellow, IEEE) received the B.S. (1985) and M.S. (1986) degrees in Electrical Engineering from Brigham Young University (BYU), and the PhD (1991) degree in Electrical Engineering from Stanford University. He was with the Department of Electrical and Computer Engineering at BYU from 1990-2007.
During 1996-97, he held a joint appointment as a visiting scholar at Uppsala University and the Royal Institute of Technology in Sweden. From 2006-07, he was on leave working as Vice President of Research for ArrayComm LLC in San Jose, California. Since 2007 he has been a Professor in the Electrical Engineering and Computer Science Department at the University of California Irvine.
During 2014-17 he was also a Hans Fischer Senior Fellow in the Institute for Advanced Studies at the Technical University of Munich. In 2016, he was elected as a Foreign Member of the Royal Swedish Academy of Engineering Sciences (IVA). His research focuses on array signal processing for radar, wireless communications, and biomedical applications, and he has over 350 publications in these areas. Dr. Swindlehurst is a Fellow of the IEEE and was the inaugural Editor-in-Chief of the IEEE Journal of Selected Topics in Signal Processing. He received the 2000 IEEE W. R. G. Baker Prize Paper Award, the 2006 IEEE Communications Society Stephen O. Rice Prize in the Field of Communication Theory, the 2006 and 2010 IEEE Signal Processing Society’s Best Paper Awards, the 2017 IEEE Signal Processing Society Donald G. Fink Overview Paper Award, and a Best Paper award at the 2020 IEEE International Conference on Communications.
\end{IEEEbiography}

\vskip -2\baselineskip plus -1fil

\begin{IEEEbiography}[{\includegraphics[width=1in,height=1.25in,clip,keepaspectratio]{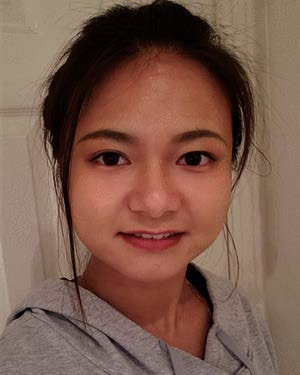}}]{Gui Zhou} (Graduate Student Member, IEEE) received the {B.S.} and {M.E.} degrees from the School of Information and Electronics, Beijing Institute of Technology, Beijing, China, in 2015 and 2019, respectively. She is currently pursuing the Ph.D. degree at the School of electronic Engineering and Computer Science, Queen Mary University of London, U.K. Her major research interests include intelligent reflection surface (IRS) and signal processing.
\end{IEEEbiography}

\vskip -2\baselineskip plus -1fil

\begin{IEEEbiography}[{\includegraphics[width=1in,height=1.25in,clip,keepaspectratio]{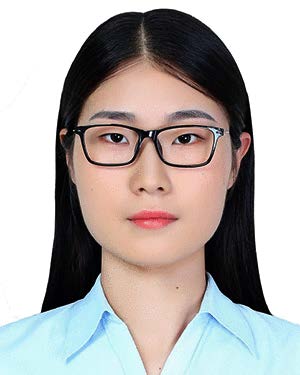}}]{Rang Liu} (Graduate Student Member, IEEE) received the B.S. degree in 2018 in electronics information engineering from the Dalian University of Technology, Dalian, China, where she is currently working toward the Ph.D. degree with the School of Information and Communication Engineering. Her current research interests include signal processing, mmWave communications, massive MIMO systems, and reconfigurable intelligent surfaces. 
\end{IEEEbiography}

\vskip -2\baselineskip plus -1fil

\begin{IEEEbiography}[{\includegraphics[width=1in,height=1.25in,clip,keepaspectratio]{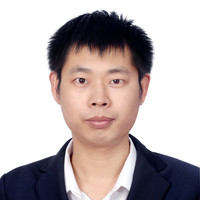}}]{Cunhua Pan} (Member, IEEE) received the B.S. and
Ph.D. degrees from the School of Information Science
and Engineering, Southeast University, Nanjing,
China, in 2010 and 2015, respectively. From 2015 to
2016, he was a Research Associate at the University of
Kent, U.K. He held a post-doctoral position at Queen
Mary University of London, U.K., from 2016 and
2019, where he is currently a Lecturer. His research
interests mainly include intelligent reflection surface
(IRS), machine learning, UAV, Internet of Things
(IoTs), and mobile edge computing. He serves as a
TPC member for numerous conferences, such as ICC and GLOBECOM, and
the Student Travel Grant Chair for ICC 2019. He also serves as an Editor of
IEEE Wireless Communication Letters, IEEE Communication Letters and IEEE Access.
\end{IEEEbiography}

\vskip -2\baselineskip plus -1fil

\begin{IEEEbiography}[{\includegraphics[width=1in,height=1.25in,clip,keepaspectratio]{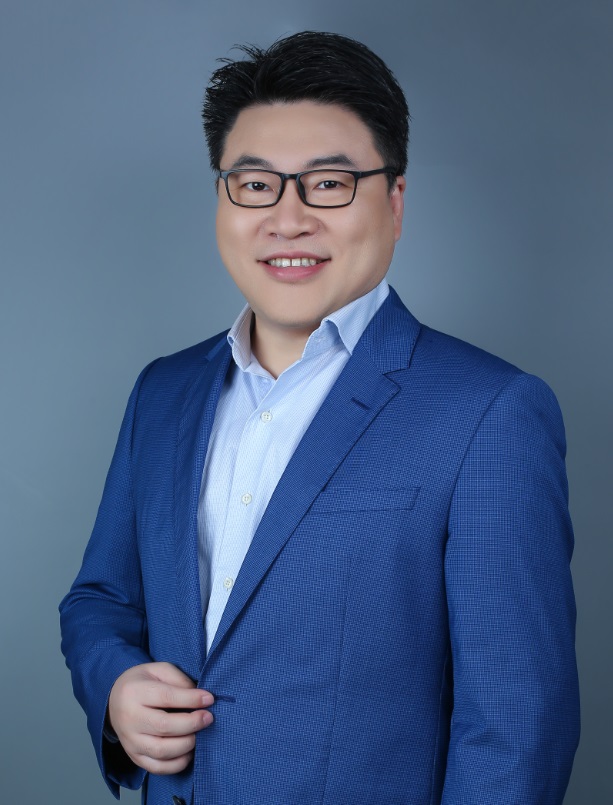}}]{Ming Li} (Senior Member, IEEE) received the
M.S. and Ph.D. degrees in Electrical Engineering
from the State University of New York at Buffalo
(SUNY-Buffalo), Buffalo, in 2005 and 2010,
respectively.
From January 2011 to August 2013, he was
a Post-Doctoral Research Associate with the Signals,
Communications, and Networking Research
Group, Department of Electrical Engineering,
SUNY-Buffalo. From August 2013 to June 2014,
he joined Qualcomm Technologies Inc., as a Senior
Engineer. Since June 2014, he has been with the School of Information and
Communication Engineering, Dalian University of Technology, Dalian, China,
where he is currently an Associate Professor. His current research interests
include the general areas of communication theory and signal processing with
applications to mmWave communications, secure wireless communications,
cognitive radios and networks, data hiding, and steganography.
\end{IEEEbiography}
\vfill
\end{document}